\documentclass[12pt,english]{article}
\usepackage[T1]{fontenc}
\usepackage[latin9]{inputenc}
\usepackage{array}
\usepackage{verbatim}
\usepackage{float}
\usepackage{rotfloat}
\usepackage{multirow}
\usepackage{amsthm}
\usepackage{amsmath}
\usepackage{amssymb}
\usepackage{graphicx}
\usepackage{setspace}
\usepackage[authoryear]{natbib}
\makeatletter

\providecommand{\tabularnewline}{\\}
\newcommand{\lyxdot}{.}

\floatstyle{ruled}
\newfloat{algorithm}{tbp}{loa}
\providecommand{\algorithmname}{Algorithm}
\floatname{algorithm}{\protect\algorithmname}

\usepackage{amsfonts}\usepackage{epsfig}\usepackage{subfig}\usepackage{amsthm}\usepackage{algorithmic}\usepackage{algorithm}

\newcommand{\Exp}{\mathbb{E}}
\theoremstyle{plain}
\newtheorem{thm}{Theorem}
\newtheorem{cor}[thm]{Corollary}

\providecommand{\abs}[1]{\lvert#1\rvert}
\providecommand{\norm}[1]{\lVert#1\rVert}

\providecommand{\dotpd}[1]{\langle#1\rangle}

\DeclareMathOperator*{\argmin}{arg\,min}
\DeclareMathOperator{\diag}{diag}
\DeclareMathOperator{\sign}{sign}

\DeclareMathOperator{\trace}{trace}
\DeclareMathOperator{\rank}{rank}
\DeclareMathOperator{\Var}{Var}

\DeclareMathOperator{\support}{support}
\DeclareMathOperator{\vech}{vec}
\DeclareMathOperator{\pen}{pen}

\def\@maketitle{%
  \newpage
  \begin{center}%
  \let \footnote \thanks
    {\LARGE \@title \par}%
    {\large
      \begin{tabular}[t]{c}%
        \@author
      \end{tabular}\par}%
    \vskip .25em%
    {\large \@date}%
  \end{center}%
  \par} 

 \renewenvironment{abstract}{%
     {\par}}

\@ifundefined{showcaptionsetup}{}{%
 \PassOptionsToPackage{caption=false}{subfig}}
\usepackage{subfig}
\makeatother

\usepackage{babel}

\begin{document}

\title{Recovering Model Structures from Large Low Rank and Sparse Covariance
Matrix Estimation %
\thanks{Xi Luo is Assistant Professor, Department of Biostatistics and Center
for Statistical Sciences, Brown University, Providence, RI 02912.
Email: xluo@stat.brown.edu.\protect \\
\indent \ \  Oral presentations of this work were given at several
departmental seminars and the joint statistical meeting, 2011. The
author would like to thank Professor Tony Cai, Dorothy Silberberg
Professor, Department of Statistics, The Wharton School, University
of Pennsylvania, for his valuable mentoring and many helpful suggestions.
An earlier version of this paper was posted and submitted in 2011
under the title ``High Dimensional Low Rank and Sparse Covariance
Matrix Estimation via Convex Minimization''. The revision of this
paper was partially supported by National Institutes of Health grants
P01-AA019072 and P30-AI042853. \protect \\
\indent \ \  An R package of the proposed method is publicly available
from the Comprehensive R Archive Network.%
}}

\author{
Xi Luo\\[-1em]Brown University
}

\maketitle

\begin{abstract}
\textbf{Abstract}. Many popular statistical models, such as factor
and random effects models, give arise a certain type of covariance
structures that is a summation of low rank and sparse matrices. This
paper introduces a penalized approximation framework to recover such
model structures from large covariance matrix estimation. We propose
an estimator based on minimizing a non-likelihood loss with separable
non-smooth penalty functions. This estimator is shown to recover exactly
the rank and sparsity patterns of these two components, and thus partially
recovers the model structures. Convergence rates under various matrix
norms are also presented. To compute this estimator, we further develop
a first-order iterative algorithm to solve a convex optimization problem
that contains separable non-smooth functions, and the algorithm is
shown to produce a solution within $O(t^{-2})$ of the optimal, after
any finite $t$ iterations. Numerical performance is illustrated using
simulated data and stock portfolio selection on S\&P 100.
\end{abstract}

\noindent \newpage{}

\section{Introduction\label{sec:intro}}

Covariance estimation is important in multivariate modeling. The natural
estimator, sample covariance, is known to perform badly in high dimensions,
where the number of variables is larger than or comparable to the
sample size. Various structural assumptions have been imposed to stabilize
this estimator. However, recovering the statistical models that lead
to such structural assumptions is under explored. As a matter of fact,
popular statistical models, such as random effects and factor models,
entail a new type of covariance structures than studied before. In
this paper, we propose a unified approach to recover the covariance
structures and the model structures from large dimensional data.

\begin{figure}[bh]
\caption{\label{fig:samplecov} Sample covariance and correlation of monthly
returns of SP100 from 1972-2009.}
\subfloat[Sample Covariance]{

\includegraphics[width=0.45\textwidth]{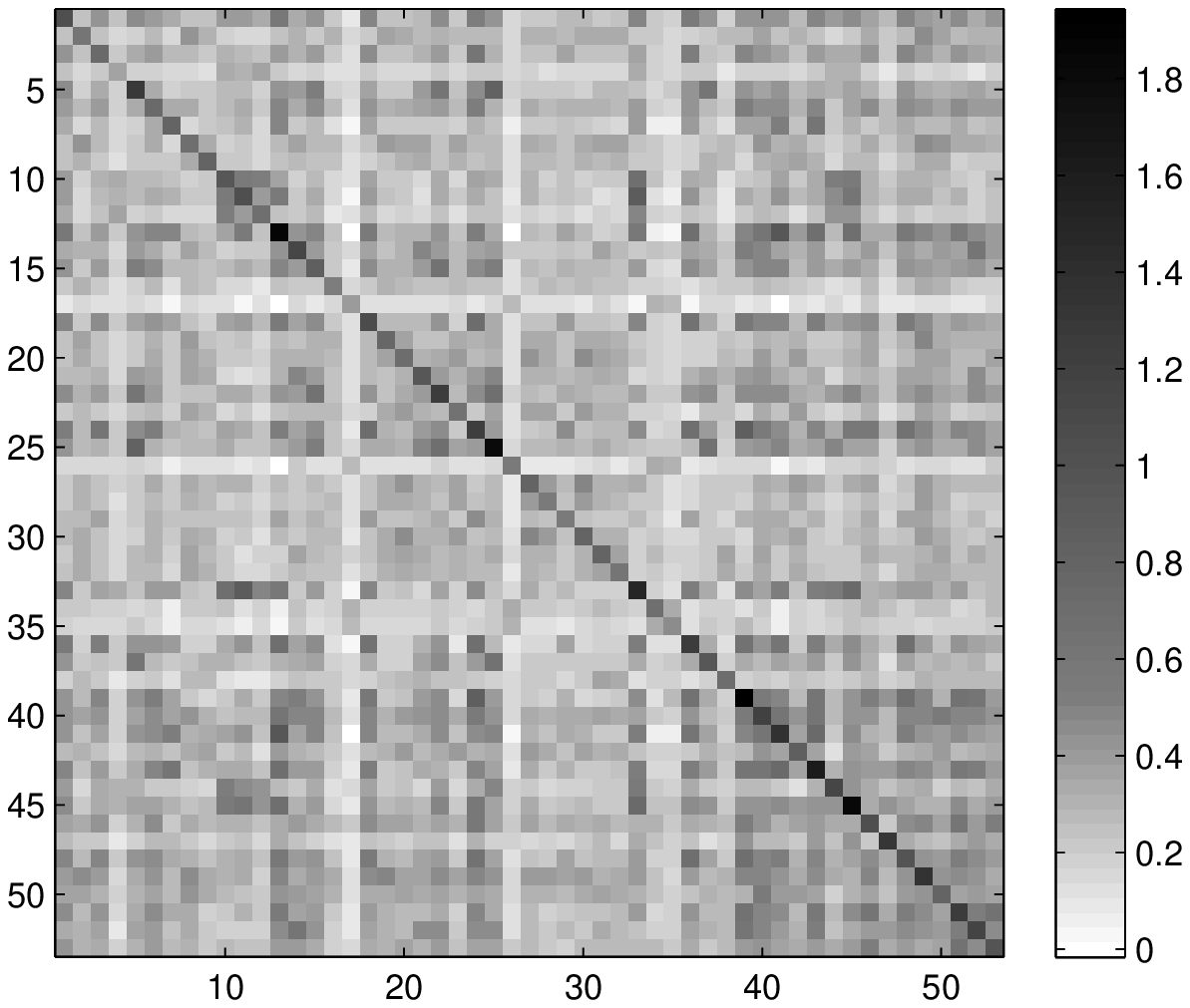}}\subfloat[Sample Correlation]{

\includegraphics[width=0.45\textwidth]{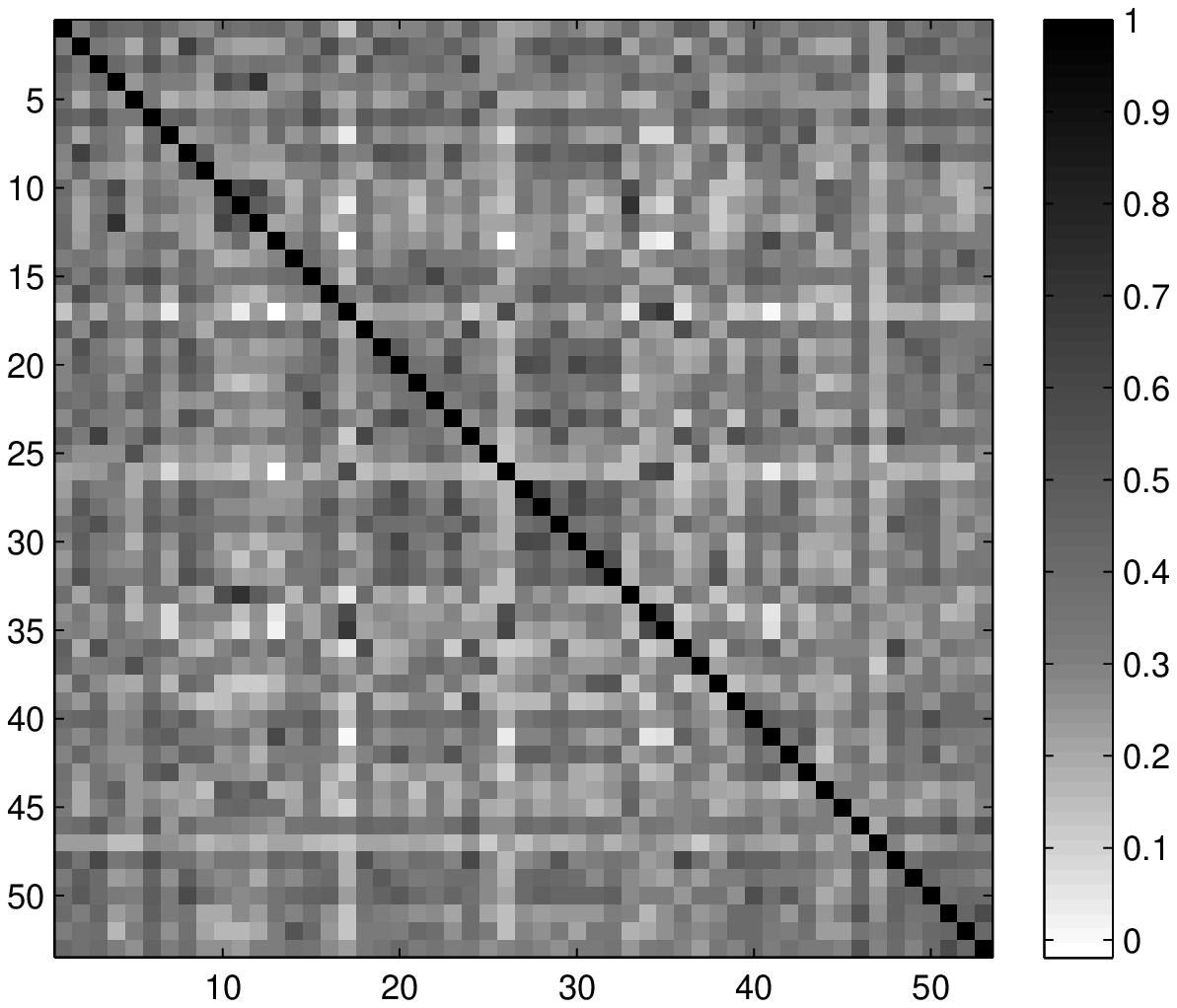}}
\end{figure}

To begin with, suppose we observe $\mathbf{X}_{1},\dotsc,\mathbf{X}_{n}\in\mathbb{R}^{p}$
iid from a multivariate Gaussian distribution $N(\mu,\mathbf{\mathbf{\mathbf{\Sigma}}}^{*})$.
The canonical estimator for $\mathbf{\Sigma}^{*}$ is the sample covariance
\[
\mathbf{\Sigma}_{n}=\frac{1}{n-1}\sum_{i=1}^{n}(\mathbf{X}_{i}-\bar{\mathbf{X}})(\mathbf{X}_{i}-\bar{\mathbf{X}})^{T}
\]
where the sample mean $\bar{\mathbf{X}}=(1/n)\sum_{i=1}^{n}\mathbf{X}_{i}$.
Figure \ref{fig:samplecov} illustrates the sample covariance and
correlation matrices of 53 monthly stock returns from year 1972-2009.
Details about this dataset can be found in Section \ref{subsec:port}.
Note that 86.14\% of the correlation entries are larger than 0.2,
more than 50\% are larger than 0.3. %

Regularizing $\mathbf{\Sigma}_{n}$ was proposed in the literature
to stabilize the estimator, mostly aiming for studying matrix norm
losses under different structural assumptions. In the time series
setting, \citet{BickelP08} proposed banding $\mathbf{\Sigma}_{n}$,
and the optimal convergence rate was established in \citet{CaiT10a}.
\citet{WuW03,HuangJ06} employed regularized regression on modified
Cholesky factors. In a setting where the indexing order is unavailable,
thresholding $\mathbf{\Sigma}_{n}$ was proposed by \citet{BickelP08b}.
Generalized thresholding rules were considered in \citet{RothmanA09},
and an adaptive thresholding rule was proposed by \citet{CaiT2011}.
There and in what follows, sparsity means the number of nonzero off-diagonal
entries is small comparing to the sample size. Without imposing such
explicit structural assumptions, shrinkage estimation was proposed
by \citet{LedoitO04}. This paper also studies the order invariant
situation, but aims to recover the model structures from covariance
estimation. First of all, it seems to be unnatural to apply thresholding
directly for our dataset in Figure \ref{fig:samplecov}, partly because
most of the entries have large magnitude. On the other hand, factor
models have been widely used for modeling stock data \citep{SharpeW64,FamaE92},
and for estimating the non-sparse covariance matrix \citep{FanJ08}.
To this end, we introduce a decomposition framework to recover a general
form of covariance structures, which includes hidden factor models
as a special case. 

This paper proposes a unified framework that entails the following
covariance structure
\begin{equation}
\mathbf{\mathbf{\Sigma}}^{*}=\mathbf{L}^{*}+\mathbf{S}^{*}\label{eq:lowspvar}
\end{equation}
where $\mathbf{L}^{*}$ is a low rank component and $\mathbf{S}^{*}$
is a sparse component. This framework is motivated by several important
statistical models. A few examples are outlined below.
\begin{enumerate}
\item Factor analysis: consider the following concrete model
\begin{equation}
\mathbf{X}=\mathbf{B}\mathbf{f}+\mathbf{\epsilon}\label{eq:facmodel}
\end{equation}
where $\mathbf{B}$ is a fixed matrix in $\mathbb{R}^{p\times K}$,
factor $\mathbf{f}\in\mathbb{R}^{K}$ and error $\mathbf{\epsilon}\in\mathbb{R}^{p}$
are independently generated from random processes. Factor model \eqref{eq:facmodel}
was widely studied in statistics \citep{AndersonT84}, in genetics
\citep{LeekJ07}, in economics and finance \citep{RossS76,ChamberlainG83,FamaE92},
and in signal processing \citep{KrimH96}. The number of factors,
$K$, is usually assumed to be small, for example, $K=1$ in the capital
asset pricing model (CAPM) \citep{SharpeW64} and $K=3$ in the Fama-French
model \citep{FamaE92}. It is easy to see that the implied covariance
matrix of model \eqref{eq:facmodel} is
\begin{equation}
\mathbf{\Sigma}^{*}=\mathbf{B}\Var(\mathbf{f})\mathbf{B}^{T}+\mathbf{\Sigma}_{\mathbf{\epsilon}}\label{eq:faccov}
\end{equation}
where the error covariance $\mathbf{\Sigma}_{\mathbf{\epsilon}}=\Exp\mathbf{\epsilon}\mathbf{\epsilon}^{T}.$
Note that $\mathbf{B}\Var(\mathbf{f})\mathbf{B}^{T}$ has at most
rank $K$. Linear regression approach under \eqref{eq:faccov} was
proposed by \citet{FanJ08}, assuming that $\mathbf{f}$ is observable
(thus $K$ is known) and $\mathbf{\Sigma}_{\mathbf{\epsilon}}$ is
diagonal. This paper instead studies a hidden $\mathbf{f}$ setting,
and we will assume a more general assumption that $\mathbf{\Sigma}_{\mathbf{\epsilon}}$
is sparse in the spirit of \citet{BickelP08b}. This more general
assumption is particularly useful in modeling stock data \citep{ChamberlainG83}.
Our method recovers both the rank $K$ and the non-zero entries in
$\mathbf{\Sigma}_{\mathbf{\epsilon}}$. An improved convergence rate
$O(n^{-1/2}(p+p^{1/2}K^{1/2})$ is also established under the Frobenius
norm, comparing with the rate $O(n^{-1/2}pK)$ in \citet{FanJ08}.
These results serve as an important step towards adapting to the order
$K$ of hidden factor models and modeling possible correlated errors.
\item Random effects: take the \emph{Compound Symmetry} model below as an
illustrating case
\begin{equation}
\mathbf{\Sigma}^{*}=\sigma_{b}^{2}\mathbf{1}^{T}\mathbf{1}+\mathbf{S}^{*}\label{eq:ComSym}
\end{equation}
where $\sigma_{b}^{2}$ and $\mathbf{S}^{*}$ are between-subject
and within-subject variances respectively. More examples with two
component decompositions similar to \eqref{eq:ComSym} can be found
a standard textbook \citep{Fitzmaurice04}. General non-diagonal $\mathbf{S}^{*}$
was studied by \citet{Jennrich86}, but only for an AR(1) type. We
aim to recover sparse $\mathbf{S}^{*}$ as well as the fist rank 1
component in \eqref{eq:ComSym}. 
\item Spiked covariance: \citet{JohnstoneI09} proposed a spiked covariance
model
\begin{equation}
\mathbf{\Sigma}^{*}=\beta\mathbf{u}\mathbf{u}^{T}+\mathbf{S}^{*}\label{eq:spikecov}
\end{equation}
where most of the coordinates of $\mathbf{u}$ are zero. Under the
assumption $\mathbf{S}^{*}=\sigma^{2}\mathbf{I}$, they proved inconsistency
results for $\mathbf{u}$ using classical PCA, and proposed a consistent
approach called \emph{sparse PCA}. \citet{AminiA09} established consistent
subset selection on the nonzero coordinates of $\mathbf{u}$, under
the scaling $n\ge Ck^{2}\log(p-k)$, where $k$ is the number of the
nonzero components in $\mathbf{u}$. Our estimator (with a simple
modification) achieves the same scaling but continues to hold for
sparse and non-diagonal $\mathbf{S}^{*}$. 
\item Conditional covariance: following \citet{AndersonT84}, suppose a
multivariate normal vector $\mathbf{X}=(\mathbf{X}_{1}^{T},\mathbf{X}_{2}^{T})^{T}$,
conditioning on the subvector $\mathbf{X}_{2}$, the conditional covariance
of the subvector $\mathbf{X}_{1}$ is
\[
\mathbf{\Sigma}^{*}=-\mathbf{\Sigma}_{12}\mathbf{\Sigma}_{22}^{-1}\mathbf{\Sigma}_{21}+\mathbf{\Sigma}_{11}
\]
where $\mathbf{\Sigma}_{ij}$, for $i,j=1,2$, is standard matrix
partition of the full covariance. The first component have rank upper
bounded by the length of $\mathbf{X}_{2}$, and the marginal covariance
$\mathbf{\Sigma}_{11}$ is assumed to be sparse. Model based approaches
have been heavily explored in the finance literature. Our method provides
an alternative approach that is non-model based and with statistical
performance guarantees. 
\end{enumerate}
In summary, our general framework \eqref{eq:lowspvar} recovers the
model structures from several statistical models. Of course, the list
of models is no way complete.

To exploit the low rank and sparse structures in \eqref{eq:lowspvar},
we propose a non-likelihood approximation criterion with a composite
penalty, called LOw Rank and sparsE Covariance estimator (LOREC).
Methods based on non-likelihood criteria were proposed by several
investigators before  \citep{YuanM10,CaiT2010,AgarwalA2011,LiuLuo2012}.
Specifically, this paper employs a Frobenius norm approximation criterion
because it enjoys several advantages in computation and theoretical
analysis. To exploit the low rank and sparse components in \eqref{eq:lowspvar},
we adopt a composite penalty combining the nuclear norm \citep{FazelM01}
and the $\ell_{1}$ norm \citep{TibshiraniR96}. The same type of
composite penalty was employed under different settings before \citep{CandesE2009,ChandrasekaranV2010}.
We extend their analysis ideas to establish the theoretical guarantees
of LOREC. Because the problems are different, LOREC achieves improved
convergence rates under a relaxed sample size requirement, see Section
\ref{sec:stat}. 

After this work was posted and submitted, a reviewer suggested a recent
work by \citet{AgarwalA2011}. There are a few major differences.
First, they studied a \emph{nearly} low-rank setting comparing with
our \emph{exact} low rank setting. Moreover, They aimed to establish
the minimax optimality under the \emph{joint squared Frobenius norm}
loss
\begin{equation}
e^{2}\left(\hat{\mathbf{L}},\hat{\mathbf{S}}\right):=\left|\hat{\mathbf{L}}-\mathbf{L}^{*}\right|_{F}^{2}+\left|\hat{\mathbf{S}}-\mathbf{S}^{*}\right|_{F}^{2}\label{eq:jointfrob}
\end{equation}
where $\hat{\mathbf{L}}$ and $\hat{\mathbf{S}}$ are the estimators
for $\mathbf{L}^{*}$ and $\mathbf{S}^{*}$ respectively. In comparison,
LOREC aims to recover model structures via the operator and max norm
bounds 
\begin{equation}
\left\Vert \hat{\mathbf{L}}-\mathbf{L}^{*}\right\Vert _{2}\quad\mbox{and}\quad\left|\hat{\mathbf{S}}-\mathbf{S}^{*}\right|_{\infty}.\label{eq:ourbounds}
\end{equation}
In Section \ref{sec:method}, we further outline several major differences
between two approaches. The structure recovery performance is further
compared and illustrated by both simulated data and a real data example
in Section \ref{sec:numerical}. Finally, this paper provides additional
algorithmic advantages which we now turn to.

To efficiently compute for large scale problems, we further develop
Nesterov's method \citep{NesterovY83,NesterovY04} to a general composite
penalization setting when two non-smooth functions are separable.
We extends the complexity analysis by \citet{JiS09} to this general
setting. Our algorithm is shown to achieve an iteration complexity
of $O(t^{-2})$, after any finite $t$ iterations. In comparison,
the finite complexity analysis was unfortunately unavailable in \citet{ChandrasekaranV2010}
and \citet{AgarwalA2011}, possibly due to the complex numerical projections
employed by their methods and algorithms. To our best knowledge, LOREC
is the first computationally trackable and publicly available algorithm
for recovering model structures from low rank and sparse covariance
matrix decompositions. 

This paper is organized as follows. In Section \ref{sec:method},
we introduce our problem set up and our method based on convex optimization.
Its statistical loss bounds are presented in Section \ref{sec:stat}.
Our algorithmic developments are discussed in Section \ref{sec:comp},
and its complexity analysis is also discussed. The numerical performance
is illustrated in Section \ref{sec:numerical}, by both simulation
examples and a real data set. Possible future directions are discussed
in Section \ref{sec:discuss}. Implications of the results and all
proofs are included in the supplementary materials online. All convergence
and complexity results are non-asymptotic.

\textbf{Notations} Let $\mathbf{M}=(M_{ij})$ be any matrix. The following
matrix norms on $\mathbf{M}$ will be used: $\abs{\mathbf{M}}_{1}=\sum_{i}\sum_{j}\abs{M_{ij}}$
stands for the elementwise $\ell_{1}$ norm; $\left\Vert \mathbf{M}\right\Vert _{1}=\max_{j}\sum_{i}\left|M_{ij}\right|$
for the matrix $\ell_{1}$ norm; $\abs{\mathbf{M}}_{\infty}=\max_{i,j}\abs{M_{ij}}$
for the elementwise $\ell_{\infty}$ norm or the max norm; $\abs{\mathbf{M}}_{F}=\sqrt{\sum_{i}\sum_{j}M_{ij}^{2}}$
for the Frobenius norm; $\norm{\mathbf{M}}_{*}=\sum_{i}D_{ii}$ for
the nuclear (trace) norm if the SVD decomposition $\mathbf{M}=\mathbf{U}\mathbf{D}\mathbf{V}^{T}$.
We denote $\dotpd{\mathbf{A},\mathbf{B}}:=\trace(\mathbf{A}^{T}\mathbf{B})$
for matrices $\mathbf{A}$ and $\mathbf{B}$ of proper sizes. We also
denote the vectorization of a matrix $M$ by $\vech(M)$. The identity
matrix is denoted by $\mathbf{I}$, and the vector of all ones is
denoted by $\mathbf{1}$. Generic constants are denoted by $C,C_{1},C_{2},\dotsc$,
and they may represent different values at each appearance.

\section{Method\label{sec:method} }

We introduce the LOREC estimator in this section. Existing approaches
that motivate LOREC will be briefly reviewed first. Other variants
and extensions of LOREC will discussed in Section \ref{sec:discuss}. 

Given a sample covariance matrix $\mathbf{\Sigma}_{n}$, low rank
approximation seeks a matrix $\mathbf{M}$ that is equivalent to optimize
\begin{equation}
\min_{\mathbf{M}}\left\{ \frac{1}{2}\left|\mathbf{M}-\mathbf{\Sigma}_{n}\right|_{F}^{2}+\lambda\rank(\mathbf{M})\right\} \label{eq:rankcon}
\end{equation}
for some penalization parameter $\lambda>0$. The multiplier $1/2$
before the Frobenius norm is not essential, and is chosen here for
simplicity in analysis. This estimator is known as low rank approximation
in matrix computation \citep{HornR94}. Recently, this rank penalty
was also studied in the regression setting \citep{BuneaF11}. Unfortunately,
its connection to model recovery was unclear, and the resulting estimator
from \eqref{eq:rankcon} is rank deficient and thus not appropriate
for estimating a full rank $\mathbf{\Sigma}^{*}$.

Recently, \citet{BickelP08b} proposed an elementwise thresholding
method. It is equivalent to the following optimization
\begin{equation}
\min_{\mathbf{M}}\left\{ \frac{1}{2}\left|\mathbf{M}-\mathbf{\mathbf{\Sigma}}_{n}\right|_{F}^{2}+\sum_{i,j}\mathrm{pen}_{\rho}(M_{ij})\right\} \label{eq:hardth}
\end{equation}
where the hard thresholding penalty is $2\mathrm{pen}_{\rho}(x)=\rho^{2}-(\left|x\right|-\rho)^{2}1\left\{ \left|x\right|<\rho\right\} $
for some thresholding level $\rho>0$. Other penalty forms were studied
in \citet{RothmanA09}. The underlying sparsity assumption in this
approach is not immediately clear under some classical multivariate
models, as illustrated in Section \ref{sec:intro}. Therefore, thresholding
provides limited model recovery. More importantly, many data examples
suggest a non-sparse covariance structure, as in Figure \ref{fig:samplecov}. 

To address these concerns, we let $\mathbf{\Sigma}^{*}$ to be a low
rank matrix plus a sparse matrix in \eqref{eq:lowspvar}. A sparse
component (possibly positive definite) overcomes the rank deficiency
of low rank approximation, and a low rank component enables model
recovery under a few statistical models discussed in Section \ref{sec:intro}.
Under the decomposition \eqref{eq:lowspvar}, it is natural to consider
a two-component estimator $\hat{\mathbf{\Sigma}}=\hat{\mathbf{L}}+\hat{\mathbf{S}}$,
with a low rank matrix $\hat{\mathbf{L}}$ and a sparse matrix $\hat{\mathbf{S}}$.
To enforce such properties, one may combine the two penalty functions
from \eqref{eq:rankcon} and \eqref{eq:hardth} in the following objective
function 
\begin{equation}
\min_{\mathbf{L},\mathbf{S}}\left\{ \frac{1}{2}\left|\mathbf{L}+\mathbf{S}-\mathbf{\Sigma}_{n}\right|_{F}^{2}+\lambda\rank(\mathbf{L})+\sum_{i,j}\mathrm{pen}_{\rho}(S_{ij})\right\} .\label{eq:rankL0}
\end{equation}
However, the objective function \eqref{eq:rankL0} is computationally
intractable, because this is a combinatorial optimization problem
in general. We replace the penalty functions with their computationally
efficient counterparts: the $\ell_{1}$ norm heuristic \citep{TibshiraniR96,FriedmanJ08},
and the nuclear norm regularization \citep{FazelM01}. We thus propose
the LOREC estimator $\hat{\mathbf{\Sigma}}=\mathbf{\hat{L}}+\hat{\mathbf{S}}$
that derives from the following optimization 
\begin{equation}
\min_{\mathbf{L},\mathbf{S}}\left\{ \frac{1}{2}\left|\mathbf{L}+\mathbf{S}-\mathbf{\Sigma}_{n}\right|_{F}^{2}+\lambda\norm{\mathbf{L}}_{*}+\rho\abs{\mathbf{S}}_{1}\right\} \label{eq:lsobj}
\end{equation}
where $\lambda>0$ and $\rho>0$ are two penalization parameters.
This objective is convex and can be solved efficiently. In Section
\ref{sec:comp}, we further develop Nesterov's algorithm to this composite
penalty setting. An important observation is that the non-smooth penalty
functions are separable in iterative updates, and thus each update
can be expressed in closed form. Complexity bounds \citep{JiS09}
are further extended to this general setting as well. Note that the
symmetry of $\left(\mathbf{L},\mathbf{S}\right)$ is always guaranteed
because the objective is invariant under transposition.

The Frobenius norm loss in \eqref{eq:lsobj} deserves some remarks.
First, it is a natural extension of the mean squared loss for vectors.
Indeed it is the $\ell_{2}$ norm of either eigenvalues or matrix
entries, and thus it is a natural distance for both low rank and sparse
matrix spaces. The second reason is that it provides a model-free
method, as covariance is also model-free. Even if a likelihood function
exists, there are some advantages of employing such non-likelihood
methods, see \citet{MeinshausenN06,CaiT2010} for example. Finally,
the Frobenius norm loss enables fast computation.

There are several variants under this framework. The sample covariance
$\mathbf{\Sigma}_{n}$ in \eqref{eq:lsobj} can be replaced by any
other reasonable estimator close to the underlying covariance. For
example, we illustrate in Section \ref{subsec:spike} by replacing
with a thresholded sample covariance. Furthermore, other penalties
can be used in \eqref{eq:lsobj} instead. For example, the diagonal
can be unpenalized by the off-diagonal $\ell_{1}$ norm defined as
$\abs{S}_{1,o}=\sum_{i\ne j}\abs{S_{ij}}$. Other possible choices
will be discussed in Section \ref{sec:discuss}. To fix ideas, we
will stick to the formulation \eqref{eq:lsobj} for the rest of the
paper.

After this paper was posted and submitted, a review suggested a recent
work by \citet{AgarwalA2011}, hereafter ANW. ANW studied theoretical
aspects using a \emph{joint} loss \eqref{eq:jointfrob} and employed
a further constrained objective function, extending \eqref{eq:lsobj}.
In our notation, their proposal can be written as 
\begin{equation}
\min_{\mathbf{L},\mathbf{S}}\left\{ \frac{1}{2}\left|\mathbf{L}+\mathbf{S}-\mathbf{\Sigma}_{n}\right|_{F}^{2}+\lambda\norm{\mathbf{L}}_{*}+\rho\abs{\mathbf{S}}_{1}\right\} \quad\mbox{such that}\,\left|\mathbf{L}\right|_{\infty}\le\frac{\alpha}{p}\label{eq:naw}
\end{equation}
where $\alpha>0$ is a third tuning parameter. Intuitively, the added
constraint on $\mathbf{L}$ introduces biases towards matrices that
are not only low rank but also have small entries, which unfortunately
is not well motivated by the statistical models discussed before.
In addition to the added complication in ANW, there are several other
major differences between ANW and LOREC, from the theoretical, computational,
and numerical prospectives. Firstly, ANW aimed to address \emph{nearly}
low-rank matrix decompositions, and the low rank component may be
biased under the additional max norm constraint. LOREC, on the other
hand, recovers exactly low rank and sparse matrices under \emph{separate}
losses \eqref{eq:ourbounds}. This separate recovery enables recovering
model structures, see Section \ref{sec:stat}. As we will also show
using extensive numerical examples in Section \ref{sec:numerical},
ANW, unfortunately, may not recover model structures while LOREC does,
as predicted by the analysis. Secondly, the Frobenius norm bounds
in ANW can be derived from the separate bounds \eqref{eq:ourbounds}
for LOREC because our bounds are stronger, though we don't establish
minimax optimality in this paper. We further argue that the joint
loss by ANW provided limited performance guarantees in many important
applications. For example, LOREC provides such guarantees for $\left(\mathbf{L}^{*}+\mathbf{S}^{*}\right)^{-1}$
under the spectral norm, which is essential for portfolio selection
on a real data example in Section \ref{subsec:port}. Thirdly, optimal
covariance estimators under the Frobenius loss were shown to be different
from the optimal ones under the spectral norm \citep{CaiT10a}. Finally,
it is unclear how the ANW estimator can be computed efficiently with
finite iteration complexity bounds, while we establish this property
in Section \ref{sec:comp}. To our best knowledge, LOREC is the first
computationally tractable approach for recovering model structures
via performing covariance matrix decomposition.

\section{Theoretical Guarantees\label{sec:stat} }

We show our main results regarding the theoretical performance of
LOREC first, without assuming any specific model structure. We then
discuss the implications under the spiked covariance model, and compare
with existing results. Implications under other models can be found
in the supplemental materials online.

\subsection{Results for Recovering General Matrix Structures \label{subsec:main}}

For the identifiability of $\mathbf{L}^{*}$ and $\mathbf{S}^{*}$,
we need some additional assumptions. These assumptions were proposed
by \citet{CandesE2009,ChandrasekaranV2010}. The following definitions
are needed for the analysis. For any matrix $\mathbf{M}\in\mathbb{R}^{p\times p}$,
define the following tangent spaces
\begin{align*}
\Omega(\mathbf{M}) & =\left\{ \mathbf{N}\in\mathbb{R}^{p\times p}\,|\,\support(\mathbf{N})\subseteq\support(\mathbf{M})\right\} ,\\
T(\mathbf{M}) & =\left\{ \mathbf{U}\mathbf{Y}_{1}^{T}+\mathbf{Y}_{2}\mathbf{V}^{T}\,|\,\mathbf{Y}_{1},\mathbf{Y}_{2}\in\mathbb{R}^{p\times r}\right\} 
\end{align*}
where the SVD decomposition $\mathbf{M}=\mathbf{U}\mathbf{D}\mathbf{V}^{T}$
with $\mathbf{U}\in\mathbb{R}^{p\times r}$, $\mathbf{V}\in\mathbb{R}^{p\times r}$
and diagonal $\mathbf{D}\in\mathbb{R}^{r\times r}$. In addition,
define the measures of coherence between $\Omega(\mathbf{M})$ and
$T(\mathbf{M})$ by 
\begin{align*}
\xi(T(\mathbf{M})) & =\max_{\mathbf{N}\in T(\mathbf{M}),\,\left\Vert \mathbf{N}\right\Vert _{2}\le1}\left|\mathbf{N}\right|_{\infty,}\\
\mu(\Omega(\mathbf{M})) & =\max_{\mathbf{N}\in\Omega(\mathbf{M}),\,\left|\mathbf{N}\right|_{\infty}\le1}\left\Vert \mathbf{N}\right\Vert _{2}.
\end{align*}
Detailed discussions of the above quantities $\Omega(\mathbf{M})$,
$T(\mathbf{M})$, $\xi(T(\mathbf{M}))$ and $\mu(\Omega(\mathbf{M}))$
and their implications can be found in \citet{ChandrasekaranV2010}.
We will give an explicit form of these quantities under the spiked
model, see Section \ref{subsec:spike}.

To characterize the convergence of $\mathbf{\Sigma}_{n}$ to $\mathbf{\Sigma}^{*}$,
we assume $\mathbf{\Sigma}^{*}$ to be within the following matrix
class
\[
\mathcal{U}(\epsilon_{0})=\left\{ \mathbf{M}\in\mathbb{R}^{p\times p}:\,0<\epsilon_{0}\le\Lambda_{i}(\mathbf{M})\le\epsilon_{0}^{-1}<+\infty,\,\mathrm{for}\, i=1,\dotsc,p\right\} 
\]
where $\Lambda_{i}(\mathbf{M})$ denotes the $i$th largest singular
value of $\mathbf{M}$. The largest eigenvalue of $\mathbf{\Sigma}^{*}$
is then $\Lambda_{\max}:=\Lambda_{1}(\mathbf{\Sigma}^{*})$, and the
smallest is $\Lambda_{\min}:=\Lambda_{p}(\mathbf{\Sigma}^{*})$. Similar
covariance classes (with additional constraints) were considered in
the bandable setting \citep{BickelP08,CaiT10a} and in the sparse
setting \citep{BickelP08b,CaiT2010,CaiT2011}. 

We have the following general theorem on the statistical performance
of LOREC. 

\begin{thm} \label{thm:loss} Let $\Omega=\Omega(\mathbf{S}^{*})$
and $T=T(\mathbf{L}^{*})$. Suppose that $\mathbf{\Sigma}^{*}\in\mathcal{U}(\epsilon_{0})$,
$\mu(\Omega)\xi(T)\le1/54$, and for $n\ge p$ that 
\[
\lambda=C_{1}\max\left(\frac{1}{\xi(T)}\sqrt{\frac{\log p}{n}},\sqrt{\frac{p}{n}}\right)
\]
and $\rho=\gamma\lambda$ where $\gamma\in\left[9\xi(T),1/(6\mu(\Omega))\right].$
In addition, suppose the minimal singular value of $\boldsymbol{L}^{*}$
is greater than $C_{2}\lambda/\xi^{2}(T)$, and the smallest absolute
value of the nonzero entries of $\mathbf{S}^{*}$ is greater than
$C_{3}\lambda/\mu(\Omega)$. Then with probability greater than $1-C_{4}p^{-C_{5}}$,
the LOREC estimator $(\hat{\mathbf{L}},\hat{\mathbf{S}})$ recovers
the rank of $\mathbf{L}^{*}$ and the sparsity pattern of $\mathbf{S}^{*}$
exactly, that is

\[
\rank(\mathbf{\hat{L}})=\rank(\mathbf{L}^{*})\quad\textrm{and}\quad\sign(\hat{\mathbf{S}})=\sign(\mathbf{S}^{*}).
\]
Moreover, with probability greater than $1-C_{4}p^{-C_{5}}$, the
matrix losses for each component are bounded as follows 
\[
\left\Vert \hat{\mathbf{L}}-\mathbf{L}^{*}\right\Vert _{2}\le C\lambda\quad\textrm{and}\quad\left|\hat{\mathbf{S}}-\mathbf{S}^{*}\right|_{\infty}\le C\rho.
\]
 \end{thm}

Theorem \ref{thm:loss} establishes the rank and sparsity recovery
results as well as the matrix loss bounds. The rank and sparsity recovery
implies that LOREC recovers the model structures from the data under
several specific models in Section \ref{sec:intro}. Under the factor
model setting, this implies that LOREC recovers the number of factors
as well as the sparsity patterns. An detailed explanation of this
merit is given in the supplemental materials online. Moreover, the
matrix loss bounds enables us to derive other loss bounds in a moment.
Note that the theoretical choice of $\lambda$ and $\rho$ are suggested
here, and we will use cross-validation to tune them in practice. 

The convergence rate for the low rank component is equivalent (up
to a $\log p$ factor) to the minimax rate in regression \citep{RohdeA11,BuneaF11,KoltchinskiiV11},
which is $O(n^{-1/2}\left(pr\right)^{1/2})$. To see it, the Frobenius
norm of $\hat{\mathbf{L}}-\mathbf{L}^{*}$ is bounded by a rate of
$O(n^{-1/2}p^{1/2}\max(r,\log p)^{1/2})$, where $\xi(T)$ above is
replaced by the lower bound $\sqrt{r/p}$ \citep{ChandrasekaranV2010}.
The additional $\log(p)$ factor in our rate is usually unavoidable
\citep{FosterD94}. We leave the optimality justification to future
research. 

The quantities $\mu(\Omega)$ and $\xi(T)$ can be explicitly determined
if certain low rank and sparse structures are assumed, for example
compound symmetry and spiked covariance. In comparison, the quantity
$\xi(T)$ plays a less important role than in the inverse covariance
problem by \citet{ChandrasekaranV2010}, because we design a new loss.
In particular, we have a smaller sample size requirement $O(p)$ than
their requirement $O(p/\xi^{4}(T))$, which could be $O(p^{3}/r^{2}$)
in the worst case. Comparing with robust PCA \citep{CandesE2009},
which studies noiseless recovery, Theorem \ref{thm:loss} establishes
the noisy recovery results as well as matrix loss bounds.

The assumption $n\ge p$ is required for technical reasons in order
to produce finite sample bounds, and the same requirement appeared
in \citet{ChandrasekaranV2010,AgarwalA2011} for example. As we show
in Section \ref{subsec:simu}, for all large $p>n$ cases, LOREC performs
almost uniformly better than any other competing method. This requirement
is due to the sample covariance used as input in \eqref{eq:lsobj}.
If a better estimate is available under stronger assumptions, this
requirement can be dropped, as we illustrate in Section \ref{subsec:spike}.
In there, a better rate is obtained due to such replacement.

\subsection{Results for Matrix Losses}

Matrix loss bounds under other norms are derived as corrollaries of
Theorem \ref{thm:loss}. 

First, the results for the joint squared Frobenius norm loss \eqref{eq:jointfrob}
in ANW is deduced as a consequence of Theorem \ref{thm:loss}. We
now compare with their results. To fix ideas, consider the scaling
$\xi\left(T\right)=O\left(\sqrt{r/p}\right)$ when the entries of
$\mathbf{L}^{*}$ are uniformly small, which is the favorable setting
for ANW. Let $s_{T}=\sum_{i,j}1\left\{ S_{ij}^{*}\ne0\right\} $ be
the total number of nonzero entries in $\mathbf{S}^{*}$. A simple
calculation yields the following bound for the LOREC estimator 
\[
e^{2}\left(\hat{\mathbf{L}},\hat{\mathbf{S}}\right)\le C\left[\frac{rp}{n}\max\left(\frac{\log p}{r},1\right)+\frac{s_{T}}{n}\max\left(\log p,r\right)\right].
\]
When the rank $r$ is the same order as $\log p$, as in the exact
low rank setting, LOREC thus achieves the same bound of ANW, see Section
3.4.1 of \citet{AgarwalA2011}. However, LOREC recovers the exact
model structures because of the rank consistency and sparsity consistency
statements in Theorem \ref{thm:loss}, whereas ANW unfortunately does
not provide such. 

As a consequence of Theorem \ref{thm:loss}, the convergence of $\mathbf{\hat{L}}+\hat{\mathbf{S}}$
to $\mathbf{\Sigma}^{*}$ can be obtained using triangular inequalities.
For instance, we show a spectral norm result after introducing additional
notations. Let $s=\max_{i}\sum_{j}1\left\{ S_{ij}^{*}\ne0\right\} $,
which is the maximum number of nonzero entries for each column. Consequently,
we have the following corollary.

\begin{cor} \label{cor:whole} Under the conditions given in Theorem
\ref{thm:loss}. Then the covariance matrix $\hat{\mathbf{L}}+\hat{\mathbf{S}}$
produced by LOREC satisfies the following spectral norm bound with
probability larger than $1-C_{1}p^{-C_{2}}$, 
\begin{equation}
\left\Vert \hat{\mathbf{L}}+\hat{\mathbf{S}}-\mathbf{\Sigma}^{*}\right\Vert \le C(s\xi(T)+1)\max\left(\frac{1}{\xi(T)}\sqrt{\frac{\log p}{n}},\sqrt{\frac{p}{n}}\right).\label{eq:specWhole}
\end{equation}
Moreover, with the same probability, the Frobenius norm bound holds
as

\begin{equation}
\left|\hat{\mathbf{L}}+\hat{\mathbf{S}}-\mathbf{\Sigma}^{*}\right|_{F}\le C(\sqrt{ps}\xi(T)+\sqrt{r})\max\left(\frac{1}{\xi(T)}\sqrt{\frac{\log p}{n}},\sqrt{\frac{p}{n}}\right).\label{eq:frobWhole}
\end{equation}

\end{cor}

The spectral norm bound \eqref{eq:specWhole} implies that the composite
estimator $\hat{\mathbf{\Sigma}}=\mathbf{\hat{L}}+\hat{\mathbf{S}}$
is positive definite if $\Lambda_{\min}$ is larger than the upper
bound there. Since the individual eigenvalue distance are bounded
by the spectral norm, see (3.5.32) from \citet{HornR94}, \eqref{eq:specWhole}
can be replaced by the following bound without any other modification
of the theorem
\begin{equation}
\max_{i}\left|\Lambda_{i}(\hat{\mathbf{L}}+\hat{\mathbf{S}})-\Lambda_{i}(\mathbf{\Sigma}^{*})\right|\le C(s\xi(T)+1)\max\left(\frac{1}{\xi(T)}\sqrt{\frac{\log p}{n}},\sqrt{\frac{p}{n}}\right):=\Phi.\label{eq:eigdistWhole}
\end{equation}
This eigenvalue convergence bound further implies the following result
concerning the inverse covariance matrix estimation. 

\begin{cor} \label{cor:inverse} Under the conditions given in Theorem
\ref{thm:loss}. Additionally, assume that $\Lambda_{\min}\ge2\Phi$
where $\Phi$ is defined in \eqref{eq:eigdistWhole}. Then the inverse
covariance matrix $(\hat{\mathbf{L}}+\hat{\mathbf{S}})^{-1}$ produced
by LOREC satisfies the following spectral norm bound with probability
larger than $1-C_{1}p^{-C_{2}}$,
\begin{equation}
\left\Vert (\hat{\mathbf{L}}+\hat{\mathbf{S}})^{-1}-(\mathbf{\Sigma}^{*})^{-1}\right\Vert \le C(s\xi(T)+1)\max\left(\frac{1}{\xi(T)}\sqrt{\frac{\log p}{n}},\sqrt{\frac{p}{n}}\right).\label{eq:inverseSpec}
\end{equation}
Moreover, with the same probability, the Frobenius norm bound holds
as
\begin{equation}
\left|(\hat{\mathbf{L}}+\hat{\mathbf{S}})^{-1}-(\mathbf{\Sigma}^{*})^{-1}\right|_{F}\le C(\sqrt{ps}\xi(T)+\sqrt{r})\max\left(\frac{1}{\xi(T)}\sqrt{\frac{\log p}{n}},\sqrt{\frac{p}{n}}\right).\label{eq:inverseFrob}
\end{equation}

\end{cor} 

This shows that the same rates hold for the inverse of the LOREC estimate
$(\hat{\mathbf{L}}+\hat{\mathbf{S}})^{-1}$. The spectral norm bound
\eqref{eq:inverseSpec} is particularly useful when the inverse covariance
is the actual input of certain methods, such as linear/quadratic discriminant
analysis, weighted least squares, and Markowitz portfolio selection.
In Section \ref{subsec:port}, we demonstrate the usage of LOREC in
portfolio selection on an S\&P 100 dataset. Again, unfortunately,
ANW does not provide the guarantees in \eqref{eq:inverseSpec}.

\subsection{Results for Spiked Covariance \label{subsec:spike}}

Recall the general spiked covariance model
\[
\mathbf{\Sigma}^{*}=\beta\mathbf{u}\mathbf{u}^{T}+\mathbf{S}^{*}
\]
where $u_{i}\ne0$ if $i\in\mathfrak{S}$ and 0 otherwise, and $\mathfrak{S}$
is a subset of $\left\{ 1,\dotsc,p\right\} $. Following \citet{AminiA09},
we fix $u_{i}\in\left\{ +1,-1\right\} /\sqrt{k}$ if $i\in\mathfrak{S}$,
and $0$ otherwise. Let the sparsity $\left|\mathfrak{S}\right|=k$.
The underlying covariance matrix is thus $\left(k+s\right)$-sparse,
i.e. at most $\left(k+s\right)$ nonzero elements for each row. \citet{BickelP08b}
showed that thresholding the sample covariance will produce an estimator
with a convergence rate $O((k+s)n^{-1/2}\log^{1/2}p)$ under the operator
norm. LOREC then achieves better convergence rates than Theorem \ref{thm:loss}
if $\mathbf{\Sigma}_{n}$ in \eqref{eq:lsobjth} is replaced by its
hard thresholded version $\mathbf{\Sigma}_{th}$ as follows
\begin{equation}
\min_{\mathbf{L},\mathbf{S}}\left\{ \frac{1}{2}\left|\mathbf{L}+\mathbf{S}-\mathbf{\Sigma}_{th}\right|_{F}^{2}+\lambda\norm{\mathbf{L}}_{*}+\rho\abs{\mathbf{S}}_{1}\right\} \label{eq:lsobjth}
\end{equation}
where $\mathbf{\Sigma}_{th}=\mathtt{T}_{\tau}\left(\mathbf{\Sigma}_{n}\right)$
with the hard thresholding rule $\mathtt{T}_{\tau}\left(M\right)=M{}_{ij}1\left\{ \left|M_{ij}\right|\ge\tau\right\} $.
The threshold $\tau$ is shown to be of the order $\sqrt{(\log p)/n}$,
and can be chosen adaptively \citep{CaiT2011}. It can be verified
that $\xi(T)\le2/\sqrt{k}$ and $\mu(\Omega)=s$ in this special case
\citep{ChandrasekaranV2010}.

\begin{cor} \label{cor:spike}

Assume the general spiked covariance model \eqref{eq:spikecov}. Suppose
that $\mathbf{\Sigma}^{*}\in\mathcal{U}(\epsilon_{0})$, $n\ge C_{0}k^{2}\log p$,
and $k\ge C_{1}s^{2}$. Let
\[
\lambda=C_{2}\left(k+s\right)\sqrt{\frac{\log p}{n}},\qquad\rho=C_{3}\left(\sqrt{k}+\sqrt{s/k}\right)\sqrt{\frac{\log p}{n}}.
\]
In addition, suppose that $\beta\ge C_{4}k\lambda$ and the smallest
absolute value of the nonzero entries of $S^{*}$ is greater than
$C_{5}\lambda$/s. Then with probability greater than $1-C_{6}p^{-C_{7}}$,
the following conclusions hold for the LOREC estimator $(\hat{L},\hat{S})$
with the input $\mathbf{\Sigma}_{th}$ in \eqref{eq:lsobjth}:
\begin{enumerate}
\item $\rank(\hat{\mathbf{L}})=1$;
\item $\sign(\hat{\mathbf{S}})=\sign(\mathbf{S}^{*})$;
\item $\left\Vert \hat{\mathbf{L}}-\beta\mathbf{u}\mathbf{u}^{T}\right\Vert \le C\left(k+s\right)n^{-1/2}(\log p)^{1/2}$;
\item $\left|\hat{\mathbf{S}}-\mathbf{S}^{*}\right|_{\infty}\le C\left(\sqrt{k}+\sqrt{s/k}\right)n^{-1/2}(\log p)^{1/2}.$
\end{enumerate}
\end{cor}

Due to the replacement $\mathbf{\Sigma}_{th}$, the assumption $n>p$
is weakened to $n\ge Ck^{2}\log p$ here, when $k$ grows not too
fast (say, smaller than $p^{\varrho}$ with $\varrho<1/2$). \citet{AminiA09}
showed that this scaling is exactly the lower bound for recovering
the support of $\mathbf{u}$ using sparse PCA \citep{JohnstoneI09}.
We will show that LOREC recovers the support under the same scaling
after a thresholding step.

To recover the support of $\mathbf{u}$, one additional thresholding
step is applied, which is a consequence of Conclusion 3 from Corollary
\ref{cor:spike}. The result is a straight forward application of
the $\sin\Theta$ theorem of Davis-Kahan \citep{DavisC70}. To see
it, the spectral bound above implies that $\left\Vert \hat{\mathbf{u}}\hat{\mathbf{u}}^{T}-\mathbf{u}\mathbf{u}^{T}\right\Vert <1/\left(2k\right)$
if $C_{4}$ is sufficiently large, where $\hat{\mathbf{u}}$ is the
eigenvector of $\hat{\mathbf{L}}$ associated with the single non-zero
eigenvalue. Therefore, thresholding $\hat{\mathbf{u}}\hat{\mathbf{u}}^{T}$
at the level $1/\left(2k\right)$ will recover the exact support of
$\mathbf{u}\mathbf{u}^{T}$, and consequently it recovers the support
of $\mathbf{u}$, with the probability stated in Corrollary \ref{cor:spike}.

\section{Algorithm and Complexity\label{sec:comp} }

We introduce an efficient algorithm for solving LOREC in this section.
Our algorithm extends Nesterov's method to a setting with separable
non-smooth penalty functions, and we then further develop iteration
complexity analysis for this setting. 

Nesterov's methods \citep{NesterovY83} have been widely used for
smooth convex optimization problems, with iteration complexity $O(1/t^{2})$
\citep{NesterovY83,NesterovY04}, where $t$ is the number of iterations.
The LOREC objective \eqref{eq:lsobjth} is, however, non-smooth, due
to two non-smooth penalty functions: the nuclear norm and the $\ell_{1}$
norm. \citet{JiS09} proposed an accelerated gradient algorithm, extending
Nesterov's method, to solve regression with a single penalty. We further
extends their work to solve the LOREC criterion that contains two
non-smooth penalty functions. One key idea that we design the LOREC
objective such that the two penalties can be separated into two optimization
problems, each with explicit updates. The analysis and algorithm are
thus extended accordingly. 

We now introduce our algorithm. Let $f(\mathbf{L},\mathbf{S})=\left|\mathbf{L}+\mathbf{S}-\mathbf{\Sigma}_{n}\right|_{F}^{2}/2$.
The gradient with respect to either $\mathbf{L}$ or $\mathbf{S}$
is of the same form 
\[
\nabla f\left(\mathbf{L},\mathbf{S}\right)=\nabla_{\mathbf{L}}f\left(\mathbf{L},\mathbf{S}\right)=\nabla_{\mathbf{S}}f\left(\mathbf{L},\mathbf{S}\right)=\mathbf{L}+\mathbf{S}-\mathbf{\Sigma}_{n}
\]
where $\nabla_{\mathbf{L}}f\left(\mathbf{L},\mathbf{S}\right)$ and
$\nabla_{\mathbf{S}}f\left(\mathbf{L},\mathbf{S}\right)$ are matrix
gradients. It is easy to see the gradient is Lipschitz continuous
with a constant $\ell=2$.

The algorithm has the following iterations. Given the previous iteration
$(\mathbf{L}_{(t-1)},\mathbf{S}_{(t-1)})$, we seek an iterative update
$(\mathbf{L}_{(t)},\mathbf{S}_{(t)})$ at iteration $t$, by minimizing
the following objective over $(\mathbf{L},\mathbf{S})$ 
\[
\begin{split}Q_{l}((\mathbf{L},\mathbf{S}),(\mathbf{L}_{(t-1)},\mathbf{S}_{(t-1)})) & =f({\mathbf{L}}_{(t-1)},{\mathbf{S}}_{(t-1)})+\left\langle \nabla_{\mathbf{L}}f_{(t-1)},\mathbf{L}-{\mathbf{L}}_{(t-1)}\right\rangle +\left\langle \nabla_{\mathbf{S}}f_{(t-1)},\mathbf{S}-{\mathbf{S}}_{(t-1)}\right\rangle \\
 & +\frac{l}{2}\abs{\mathbf{L}-\mathbf{L}_{(t-1)}}_{F}^{2}+\frac{l}{2}\abs{\mathbf{S}-\mathbf{S}_{(t-1)}}_{F}^{2}+\lambda\norm{\mathbf{L}}_{*}+\rho\abs{\mathbf{S}}_{1}
\end{split}
\]
where $(\nabla_{\mathbf{L}}f_{(t)},\nabla_{\mathbf{S}}f_{(t)}):=(\nabla_{\mathbf{L}}f(\mathbf{L}_{(t)},\mathbf{S}_{(t)}),\nabla_{\mathbf{S}}f(\mathbf{L}_{(t)},\mathbf{S}_{(t)}))$.
A similar update rule with a single non-smooth penalty was proposed
by \citet{JiS09}. Instead of adaptively searching for the stepsize
$l$, one can simply choose a fixed $l\ge\ell=2$, as we will do in
this paper. The initializer $\left(\mathbf{L}_{(0)},\mathbf{S}_{(0)}\right)$
can be simply set to be $\left(\diag\left(\mathbf{\Sigma}_{n}\right),\diag\left(\mathbf{\Sigma}_{n}\right)\right)/2$. 

A key observation that the objective $Q_{l}$ is separable in $\mathbf{L}$
and $\mathbf{S}$. Thus the optimization over $Q_{l}$ is equivalent
to seek $\mathbf{L}$ and $\mathbf{S}$ respectively in the following
two optimization problems: 
\begin{align}
\mathbf{L}_{(t)} & =\argmin_{\mathbf{L}}\{\frac{l}{2}\abs{\mathbf{L}-(\mathbf{L}_{(t-1)}-\frac{1}{l}\nabla f_{(t-1)})}_{F}^{2}+\lambda\norm{\mathbf{L}}_{*}\}\label{eq:lstep}\\
\mathbf{S}_{(t)} & =\argmin_{\mathbf{S}}\{\frac{l}{2}\abs{\mathbf{S}-(\mathbf{S}_{(t-1)}-\frac{1}{l}\nabla f_{(t-1)})}_{F}^{2}+\rho\abs{\mathbf{S}}_{1}\}.\label{eq:sstep}
\end{align}
The two objectives can be solved explicitly by soft-thresholding the
singular values and matrix entries of $\mathbf{L}_{(t-1)}-\frac{1}{l}\nabla f_{(t-1)}$
and $\mathbf{S}_{(t-1)}-\frac{1}{l}\nabla f_{(t-1)}$ respectively
\citep{JiS09,CaiJ10,FriedmanJ08}. The soft-thresholding operator
is denoted by $\mathcal{T}_{\lambda}\left(M\right)=\sign\left(M\right)\max\left(\left|M\right|-\lambda,0\right)$.
Additional algorithmic derivation can be found in the supplemental
materials online. 

After obtaining the updates, our final algorithm applies Nesterov's
mixing to consecutive updates \citep{NesterovY83}, which is summarized
in Algorithm \ref{alg:grad2}. 
\begin{algorithm}
\caption{Algorithm for LOREC.}

\begin{algorithmic} \STATE Initialize $\mathbf{L}=\mathbf{L}_{(0)}=\mathbf{Y}_{(1)}$,
and $\mathbf{S}=\mathbf{S}_{(0)}=\mathbf{Z}_{(1)}$. \REPEAT \STATE
Set $\mathbf{L}_{(t)}=\mathbf{U}\mathbf{D}_{\lambda/l}\mathbf{V}^{T}$
where the SVD $\mathbf{Y}_{(t-1)}-\frac{1}{l}\nabla f_{(t-1)}\left(\mathbf{Y}_{(t-1)},\mathbf{Z}_{(t-1)}\right)=\mathbf{U}\mathbf{D}\mathbf{V}^{T}$
and $\mathbf{D}_{\lambda/l}=\diag\left(\mathcal{T}_{\lambda/l}\left(D_{11}\right),\dotsc,\mathcal{T}_{\lambda/l}\left(D_{pp}\right)\right)$$ $.
\STATE Set $\mathbf{S}_{(t)}=\left(\mathcal{T}_{\rho/l}\left(M_{ij}\right)\right)$
where $\mathbf{M}={\mathbf{Z}}_{(t-1)}-\frac{1}{l}\nabla f_{(t-1)}\left(\mathbf{Y}_{(t-1)},\mathbf{Z}_{(t-1)}\right)$.
\STATE Set $(\mathbf{Y}_{t+1},\mathbf{Z}_{(t+1)})=(\mathbf{L}_{(t)},\mathbf{S}_{(t)})+(\frac{\alpha_{t}-1}{\alpha_{t+1}})[(\mathbf{L}_{(t)},\mathbf{S}_{(t)})-(\mathbf{L}_{(t-1)},\mathbf{S}_{(t-1)})]$
where $\alpha_{t+1}=\frac{1+\sqrt{1+4\alpha_{t}^{2}}}{2}$. \UNTIL
Convergence criterion: $\frac{\left|\mathbf{L}_{(t)}-\mathbf{L}_{(t-1)}\right|_{F}}{1+\left|\mathbf{L}_{(t-1)}\right|_{F}}+\frac{\left|\mathbf{S}_{(t)}-\mathbf{S}_{(t-1)}\right|_{F}}{1+\left|\mathbf{S}_{(t-1)}\right|_{F}}\le\epsilon$.
\end{algorithmic} \label{alg:grad2} 
\end{algorithm}
 We establish the global convergence rate of Algorithm~\ref{alg:grad2},
extending the ideas in \citet{BeckA09,JiS09} to our setting with
separable penalties.

\begin{thm} \label{thm:congrad2} Let $(\mathbf{L}_{(t)},\mathbf{S}_{(t)})$
be the update produced by Algorithm~\ref{alg:grad2} at iteration
$t$. Then for any $t\ge1$, we have the following computational accuracy
bound 
\[
F(\mathbf{L}_{(t)},\mathbf{S}_{(t)})-F(\hat{\mathbf{L}},\hat{\mathbf{S}})\le8\frac{\abs{\mathbf{L}_{(0)}-\hat{\mathbf{L}}}_{F}^{2}+\abs{\mathbf{S}_{(0)}-\hat{\mathbf{S}}}_{F}^{2}}{(t+1)^{2}}
\]
where $\left(\hat{\mathbf{L}},\hat{\mathbf{S}}\right)$ is the minimizer
of the LOREC objective $F\left(\mathbf{L},\mathbf{S}\right)=f\left(\mathbf{L},\mathbf{S}\right)+\lambda\left\Vert \mathbf{L}\right\Vert _{*}+\rho\left|\mathbf{S}\right|_{1}$
. \end{thm} 

It is easy to see the memory cost for this algorithm is $O\left(p^{2}\right)$
which is the minimal requirement for storing a general covariance
matrix. Algorithm \ref{alg:grad2} needs at most $O(p^{4}/\sqrt{\epsilon})$
computational operations to achieve an computational accuracy of $\epsilon$,
which is established as follows. Simply, the number of operations
at each iteration is $O(p^{3})$ due to full SVD, which is of the
same order of least squares. The numerator in the bound above is at
most $O(p^{2})$, and the crude bound $O(p^{4}/\sqrt{\epsilon})$
for an $\epsilon$-optimal solution is immediate. This is a crude
bound as one can be easily improve with partial SVD as discussed in
Section \ref{sec:discuss}. For a comparison, linear programming for
$O(p^{2})$ variables takes $O(p^{6}/\log(\epsilon))$ operations
using interior point methods, as in \citet{CaiT2010} for the inverse
covariance problem. Algorithm \ref{alg:grad2} is $O(p^{2})$ faster
than linear programming if the precision requirement $\epsilon$ is
not high.

\section{Numerical Results\label{sec:numerical}}

In this section, we compare the performance of LOREC with other methods
using simulated data and a real data example.

\subsection{Simulation\label{subsec:simu}}

The data are simulated from the following covariance models:
\begin{enumerate}
\item \textbf{factor}: $\mathbf{\Sigma}^{*}=\mathbf{U}\mathbf{D}\mathbf{U}^{T}+\mathbf{I}$
where $\mathbf{U}\in\mathbb{R}^{p\times3}$ with uniform orthonormal
columns, and $\mathbf{D}=\diag(8,8,8)$.
\item \textbf{compound symmetry}: random permutation of $\Sigma^{*}=0.2\mathbf{1}\mathbf{1}^{T}+\mathbf{S}$,
where $\mathbf{S}=\diag\left(\mathbf{B},\mathbf{B},\dotsc,\mathbf{B}\right)$
is block diagonal with each square block matrix $\mathbf{B}$ of dimension
5, and $\mathbf{B}=0.4\mathbf{1}\mathbf{1}^{T}+0.4\mathbf{I}$. 
\item \textbf{spike}: $\mathbf{\Sigma}^{*}=\beta\mathbf{u}\mathbf{u}^{T}+\mathbf{S}$
with $\beta=16$, $\left|\left\{ j:\, u_{j}\ne0\right\} \right|=p/2$,
and $\mathbf{S}$ is a block diagonal matrix. Each block in $\mathbf{S}$
is $0.4\mathbf{1}\mathbf{1}^{T}+0.6\mathbf{I}$ of size $4$. 
\end{enumerate}
We compare the performance of LOREC with the sample covariance, thresholding
\citep{BickelP08b}, Shrinkage \citep{LedoitO04} and the ANW estimator
\citep{AgarwalA2011}. 

For each model, $100$ observations are generated from multivariate
Gaussian distribution. The tuning parameters for the LOREC, thresholding,
shrinkage and ANW estimators are picked by $5$-fold cross validation
using the Frobenius loss. \citet{AgarwalA2011} suggested a theoretically
tuned ANW estimator where the tuning parameters are set explicitly
from knowing the underly simulation models. However we found this
theoretical choice usually performs worse than cross validation. Thus
we only include the cross validated ANW estimator in the following
comparison. The sample covariance estimator does not have a tuning
parameter, and is calculated using the whole $100$ samples. The process
is replicated $100$ times with varying $p=120,200,400$. 

Table \ref{tab:simu2-spec}-\ref{tab:simu2-frob} compare the matrix
loss performance measured by the operator norm and the Frobenius norm
respectively. The LOREC estimator almost uniformly outperforms all
other competing methods, except for the compound symmetry model where
LOREC is the second best after ANW. However, as we show in Table \ref{tab:ranksupportnew},
ANW (as expected from their theory) systematically overestimate the
ranks of the low rank components, because it biases towards a nearly
low-rank and small-entry setting. It provides little information on
recovering the model structures, even though it may have improvement
in matrix losses under this special case.

\begin{table}
\caption{\label{tab:simu2-spec} Comparison of average (SE) spectral losses
of LOREC, sample, thresholding and ANW estimators. The best performance
is shown in bold.}

\begin{tabular}{c|ccccc}
\multicolumn{6}{c}{Factor}\tabularnewline
\hline 
p & LOREC & Sample & Threshold & Shrinkage & ANW\tabularnewline
\hline 
120 & \textbf{4.91(0.06)} & 5.42(0.07) & 7.35(0.06) & 5.63(0.05) & 5.06(0.06)\tabularnewline
200 & \textbf{5.94(0.06)} & 6.91(0.08) & 7.84(0.003) & 6.71(0.04) & 6.05(0.05)\tabularnewline
400 & \textbf{7.16(0.05)} & 10.33(0.07) & 7.92(0.001) & 7.57(0.006) & 7.30(0.04)\tabularnewline
\hline 
\multicolumn{6}{c}{}\tabularnewline
\multicolumn{6}{c}{Compound Symmetry}\tabularnewline
\hline 
p & LOREC & Sample & Threshold & Shrinkage & ANW\tabularnewline
\hline 
120 & 7.50(0.21) & 7.15(0.18) & 7.88(0.18) & 7.87(0.23) & \textbf{5.96(0.22)}\tabularnewline
200 & 11.93(0.33) & 11.26(0.23) & 12.37(0.32) & 13.46(0.42) & \textbf{9.87(0.38)}\tabularnewline
400 & 23.62(0.65) & 22.00(0.58) & 24.73(0.68) & 27.44(0.79) & \textbf{19.02(0.72)}\tabularnewline
\hline 
\multicolumn{6}{c}{}\tabularnewline
\multicolumn{6}{c}{Spike}\tabularnewline
\hline 
p & LOREC & Sample & Threshold & Shrinkage & ANW\tabularnewline
\hline 
120 & 5.88(0.12) & 5.98(0.12) & 6.87(0.18) & 7.75(0.16) & \textbf{5.84(0.13)}\tabularnewline
200 & \textbf{7.20(0.16)} & 7.82(0.13) & 14.30(0.14) & 10.38(0.15) & 7.40(0.14)\tabularnewline
400 & \textbf{9.80(0.17)} & 11.36(0.14) & 15.48(0.02) & 13.43(0.05) & 9.94(0.14)\tabularnewline
\hline 
\end{tabular}
\end{table}
 
\begin{table}
\caption{\label{tab:simu2-frob} Comparison of average (SE) Frobenius losses
of LOREC, sample, thresholding and ANW estimators. The best performance
is shown in bold.}

\begin{tabular}{c|ccccc}
\multicolumn{6}{c}{Factor}\tabularnewline
\hline 
p & LOREC & Sample & Threshold & Shrinkage & ANW\tabularnewline
\hline 
120 & \textbf{7.99(0.05)} & 14.63(0.04) & 13.56(0.02) & 10.14(0.05) & 8.49(0.05)\tabularnewline
200 & \textbf{9.62(0.05)} & 22.58(0.04) & 13.87(0.002) & 11.94(0.04) & 10.36(0.05)\tabularnewline
400 & \textbf{11.81(0.04)} & 42.72(0.04) & 14.10(0.002) & 13.21(0.01) & 13.17(0.04)\tabularnewline
\hline 
\multicolumn{6}{c}{}\tabularnewline
\multicolumn{6}{c}{Compound Symmetry}\tabularnewline
\hline 
p & LOREC & Sample & Threshold & Shrinkage & ANW\tabularnewline
\hline 
120 & 9.17(0.15) & 12.49(0.09) & 13.89(0.10) & 11.46(0.14) & \textbf{7.83(0.16)}\tabularnewline
200 & 14.35(0.24) & 20.48(0.11) & 22.91(0.17) & 19.13(0.26) & \textbf{12.27(0.29)}\tabularnewline
400 & 27.55(0.51) & 41.01(0.34) & 45.91(0.41) & 38.13(0.47) & \textbf{22.37(0.60)}\tabularnewline
\hline 
\multicolumn{6}{c}{}\tabularnewline
\multicolumn{6}{c}{Spike}\tabularnewline
\hline 
p & LOREC & Sample & Threshold & Shrinkage & ANW\tabularnewline
\hline 
120 & \textbf{8.24(0.07)} & 13.85(0.06) & 14.26(0.06) & 11.03(0.07) & 8.26(0.08)\tabularnewline
200 & \textbf{10.47(0.09)} & 21.87(0.05) & 18.40(0.06) & 14.31(0.07) & 10.69(0.08)\tabularnewline
400 & \textbf{14.56(0.10)} & 41.92(0.05) & 20.25(0.02) & 18.98(0.03) & 15.16(0.07)\tabularnewline
\hline 
\end{tabular}
\end{table}

Both LOREC and ANW estimators produce estimates for the rank and the
sparsity patterns. Table \ref{tab:ranksupportnew} compares the frequencies
of exact rank and support recovery for LOREC and ANW. The singular
values and entry values are considered to be nonzero if their magnitudes
exceed $10^{-3}$ in both methods, since the error tolerance is set
to $10^{-4}$. Note that the true rank of the low rank components
in these three models are 3, 1, and 4 respectively. It is clear that
LOREC recovers both the rank and sparsity pattern with high frequencies
for all the three models, as predicted by our theory. ANW recovers
the sparsity patterns, however, it almost always over estimates the
number of ranks, resulting only nearly low rank components (rank>10
in almost all cases), which is expected from the aims of ANW. As a
matter of fact, nearly low rank simulation models with rank between
10 and 50 were reported in \citet{AgarwalA2011}, which is consistent
with our simulation findings.

\begin{sidewaystable}
\caption{\label{tab:ranksupportnew}Comparison of rank recovery for the low
rank component, and support recovery for the sparse component, by
LOREC and ANW, over 100 runs. The best performance is shown in bold. }

\begin{centering}
\begin{tabular}{cccccccccc}
\hline 
\multicolumn{10}{c}{Low Rank Recovery}\tabularnewline
\hline 
 &  & \multicolumn{2}{c}{p=120} &  & \multicolumn{2}{c}{p=200} &  & \multicolumn{2}{c}{p=400}\tabularnewline
\cline{3-4} \cline{6-7} \cline{9-10} 
 &  & LOREC & ANW &  & LOREC & ANW &  & LOREC & ANW\tabularnewline
\hline 
\multirow{2}{*}{Factor} & \%(rank=3) & \textbf{80} & 4 &  & \textbf{99} & 1 &  & \textbf{91} & 0\tabularnewline
 & mean(se) & \textbf{3.42(0.09)} & 8.66(0.34) &  & \textbf{3.01(0.01)} & 15.32(0.53) &  & \textbf{2.91(0.03)} & 25.46(0.64)\tabularnewline
\hline 
Compound & \%(rank=1) & \textbf{71} & 0 &  & \textbf{82} & 0 &  & \textbf{80} & 0\tabularnewline
Symmetry & mean(se) & \textbf{1.52(0.09)} & 60.44(0.91) &  & \textbf{1.46(0.11)} & 96.51(1.54) &  & \textbf{1.56(0.13)} & 189.84(3.54)\tabularnewline
\hline 
\multirow{2}{*}{Spike} & \%(rank=1) & \textbf{69} & 0 &  & \textbf{78} & 0 &  & \textbf{95} & 0\tabularnewline
 & mean(se) & \textbf{1.42(0.08)} & 16.86(0.66) &  & \textbf{1.22(0.04)} & 15.04(0.61) &  & \textbf{1.01(0.02)} & 11.37(0.60)\tabularnewline
\hline 
\end{tabular}
\par\end{centering}

\vspace{4em}

\centering{}%
\begin{tabular}{cccccccccc}
\hline 
\multicolumn{10}{c}{Sparse Support Recovery}\tabularnewline
\hline 
 &  & \multicolumn{2}{c}{p=120} &  & \multicolumn{2}{c}{p=200} &  & \multicolumn{2}{c}{p=400}\tabularnewline
\cline{3-4} \cline{6-7} \cline{9-10} 
 &  & LOREC & ANW &  & LOREC & ANW &  & LOREC & ANW\tabularnewline
\hline 
\multirow{2}{*}{Factor} & \%TP & \textbf{100.0(0.0)} & \textbf{100.0(0.0)} &  & \textbf{100.0(0.0)} & \textbf{100.0(0.0)} &  & \textbf{100.0(0.0)} & \textbf{100.0(0.0)}\tabularnewline
 & \%TN & \textbf{99.97(0.01)} & 95.95(0.11) &  & \textbf{100.0(0.001)} & 97.91(0.11) &  & \textbf{100.0(0.0003)} & 99.41(0.004)\tabularnewline
\hline 
Compound & \%TP & \textbf{99.95(0.02)} & 99.69(0.04) &  & \textbf{99.91(0.02)} & 99.63(0.04) &  & \textbf{99.81(0.02)} & 99.20(0.06)\tabularnewline
Symmetry & \%TN & \textbf{91.96(0.29)} & 91.90(0.23) &  & \textbf{94.76(0.18)} & 94.37(0.18) &  & \textbf{97.13(0.09)} & 97.11(0.09)\tabularnewline
\hline 
\multirow{2}{*}{Spike} & \%TP & 97.59(0.19) & \textbf{98.05(0.08)} &  & 97.04(0.20) & \textbf{97.65(0.09)} &  & 94.28(0.30) & \textbf{94.44(0.15)}\tabularnewline
 & \%TN & \textbf{92.31(0.18)} & 92.05(0.10) &  & \textbf{94.42(0.14)} & 93.76(0.07) &  & 97.57(0.14) & \textbf{97.93(0.09)}\tabularnewline
\hline 
\end{tabular}
\end{sidewaystable}

\subsection{Application to Portfolio Selection\label{subsec:port}}

We now compare performance when applying covariance estimation to
portfolio selection. \citet{MarkowitzH52} suggested constructing
the minimal variance portfolio $\mathbf{w}$ by the following optimization
\[
\min\;\mathbf{w}^{T}\mathbf{\Sigma}\mathbf{w}\quad\textrm{subject to:}\;\mathbf{w}^{T}\mathbf{1}=1\;\textrm{and}\;\mathbf{w}^{T}\mathbf{\mu}=q
\]
where $\mathbf{\mu}$ is the expected return vector and $q$ is the
required expected return of the portfolio. The solution of the problem
has the following form

\begin{equation}
\mathbf{w}=\frac{A_{3}-qA_{2}}{A_{1}A_{3}-A_{2}^{2}}\mathbf{\Sigma}^{-1}\mathbf{1}+\frac{qA_{1}-A_{2}}{A_{1}A_{3}-A_{2}^{2}}\mathbf{\Sigma}^{-1}\mathbf{\mu}\label{eq:port}
\end{equation}
where $A_{1}=\mathbf{1}^{T}\mathbf{\Sigma}\mathbf{1}$, $A_{2}=\mathbf{1}^{T}\mathbf{\Sigma}\mu$,
and $A_{3}=\mathbf{\mu}^{T}\mathbf{\Sigma}\mathbf{\mu}$. The global
minimal variance return without constraining $q$ is obtained by replacing
$q=A_{2}/A_{3}$. The portfolio \eqref{eq:port} is constructed using
an estimated covariance $\mathbf{\Sigma}$, and we compare the performance
from various covariance estimators. In addition to the covariance
matrix estimators considered in the simulations before, we include
another shrinkage estimator from \citet{LedoitO03}: ``shrink to
market''. The details of this estimator can be found therein. %

The monthly returns of stocks listed in S\&P 100 (as of December 2010)
are extracted from Center for Research in Security Prices for the
period from January 1972 to December 2009. We remove the stocks with
missing monthly returns within the extraction period due to company
history or reorganization, and 53 stocks are retained for the following
analysis. All monthly returns are annualized by multiplying 12. The
sample covariance and correlation are plotted in Figure \ref{fig:samplecov}.
Most of the entries are moderate. Similar observations (not shown)
are drawn after removing risk-free returns.

We conduct the following strategy in building our portfolios and evaluate
their forecasting performance, similar to \citet{LedoitO03}. On the
beginning of testing year $y$, we estimate the covariance matrix
from the past 10 years from January of year $y-10$ to December of
year $y-1$. We then construct the portfolio using \eqref{eq:port},
and hold it throughout year $y$. The monthly returns of the resulting
portfolio during year $y$ are recorded. In thresholding, shrinkage,
ANW, and LOREC, we pick the parameters that produce the smallest overall
out-of-sample variance for each testing year from $y-5$ to $y-1$
by constructing portfolios using their past 10 year covariance estimates. 

Table \ref{tab:port} compares the variances of realized returns during
year 1987-2009, where the covariance in the portfolio weight \eqref{eq:port}
is estimated from preceding 10-year data by different methods. LOREC
almost outperforms all other competing methods. It is slightly worse
by a nominal amount than ``shrink to market'' when the restriction
$q$ is set to be $10\%$. The ANW estimator performs almost as good
as LOREC in these examples, however, ANW is not expected to recover
the underlying model structures as we illustrate next. %

\begin{table}
\caption{\label{tab:port} Comparison of average (se) of realized return variances
for year 1987-2009. The best performance is shown in bold. }

\centering{}%
\begin{tabular}{|c|c|c|c|c|}
\hline 
 & Unconstrained & $10\%$ Constr. & $15\%$ Constr. & $20\%$ Constr.\tabularnewline
\hline 
\hline 
Sample & 0.379(0.049) & 0.403(0.054) & 0.370(0.047) & 0.377(0.046)\tabularnewline
\hline 
Thresholding & 0.376(0.047) & 0.400(0.052) & 0.365(0.045) & 0.371(0.044)\tabularnewline
\hline 
Shrink to identity & 0.284(0.035) & 0.290(0.037) & 0.263(0.033) & 0.265(0.034)\tabularnewline
\hline 
Shrink to market & 0.249(0.029) & \textbf{0.272(0.034)} & 0.249(0.030) & 0.251(0.031)\tabularnewline
\hline 
ANW & 0.249(0.032) & 0.277(0.037) & 0.250(0.037) & 0.241(0.045)\tabularnewline
\hline 
LOREC & \textbf{0.227(0.038)} & 0.278(0.048) & \textbf{0.236(0.041)} & \textbf{0.221(0.038)}\tabularnewline
\hline 
\end{tabular}
\end{table}

Both LOREC and ANW provide the decomposition of the low rank and sparse
components in stock data. LOREC identifies a rank one component, for
each 10-year periods ending from 2001 to 2009, but none for the earlier
years. This single rank finding is consistent with a rank test result
on a similar dataset \citep{OnatskiA09}. In comparison, consistent
with the simulation findings in \citet{AgarwalA2011} and ours, ANW
tends to over estimate the number of factors. Across the same range
of years, the low rank component by ANW has an average rank 3.7 (sd=5.3)
for unconstrained portfolio, 3.2 (sd=4.3) for 10\% constrained portfolios,
3 (sd=4.3) for 15\%, and 3 (sd=4.3) for 20\%. It is also worthy to
note that the standard deviations of these ANW ranks are comparatively
large, indicating unstable rank recovery. Again, one would not expect
ANW to recover the exact number of ranks (thus the number of factors),
and it is thus not fair to further compare the model recovery performance
of ANW in what follows. 

We now compare the recovered low rank component in LOREC with a single
factor model, Capital Asset Pricing Model (CAPM) \citep{SharpeW64}.
The loading (also called beta's) can be estimated using additional
proxies. Here the proxy is the usual choice of market returns. The
singular vector of an rank 1 component is equivalent (up to a multiplying
constant) to the loading in a single factor model (see supplemental
materials online). In Figure \ref{fig:angle}, we compare the loading
estimates from LOREC and CAPM by plotting the cosine of the angles
between these two vectors for year 2001-2009. The plot shows that
the angle of these two loading vectors is close to 0 consistently,
almost monotone decreasing from $27.33$ to $9.90$ degrees. It suggests
that LOREC verifies a similar loading as CAPM, though it does not
need to employ a proxy. It is also interesting to notice the trend
that the angles between these two estimates become smaller as time
progresses.

\begin{figure}
\caption{\label{fig:angle}Correlations (left panel) and angles (right panel)
between the loading vectors recovered by LOREC and CAPM.}

\begin{minipage}[t]{0.45\textwidth}%
\includegraphics[scale=0.45]{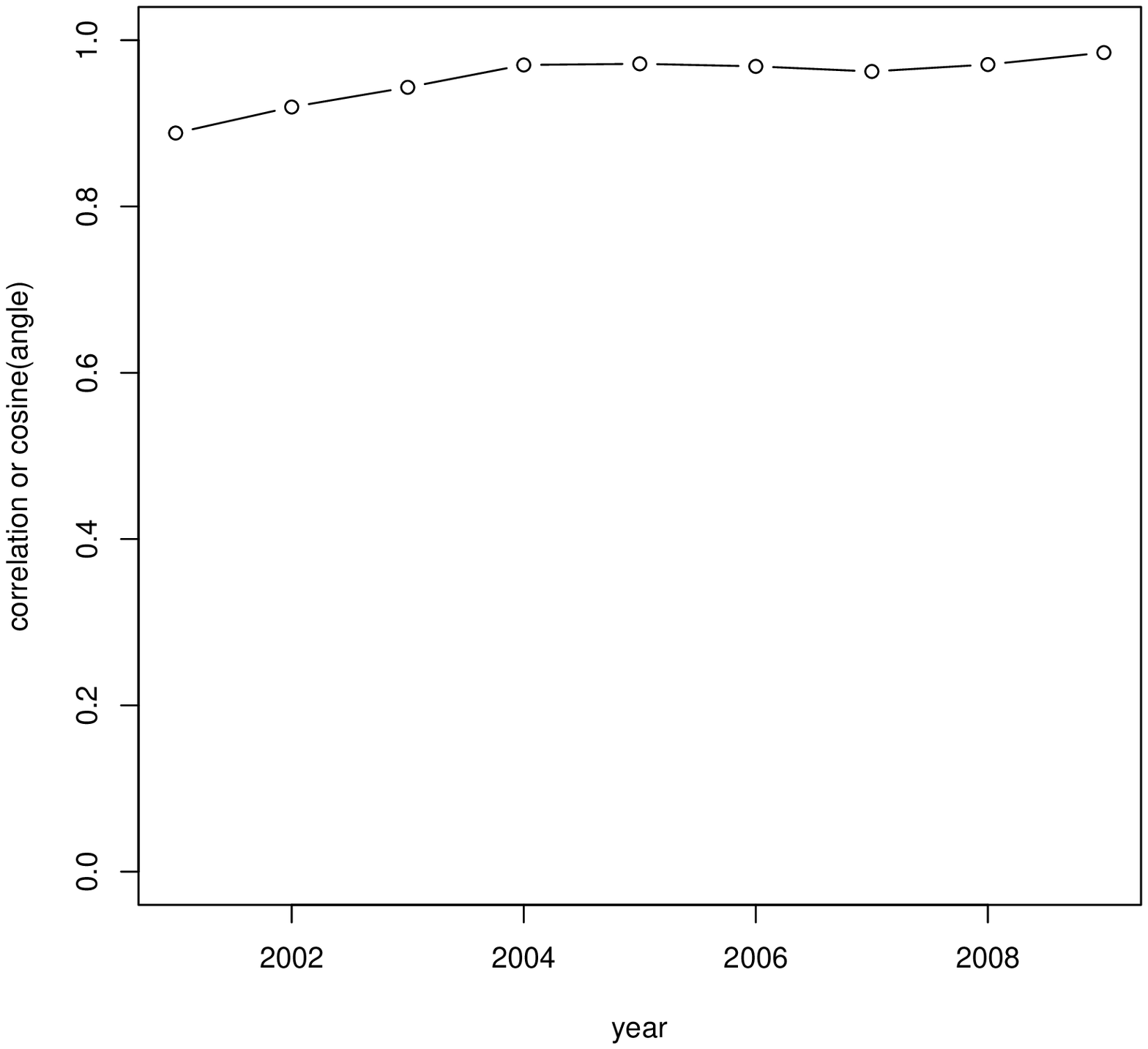}%
\end{minipage}\hfill{}%
\begin{minipage}[t]{0.45\textwidth}%
\includegraphics[scale=0.45]{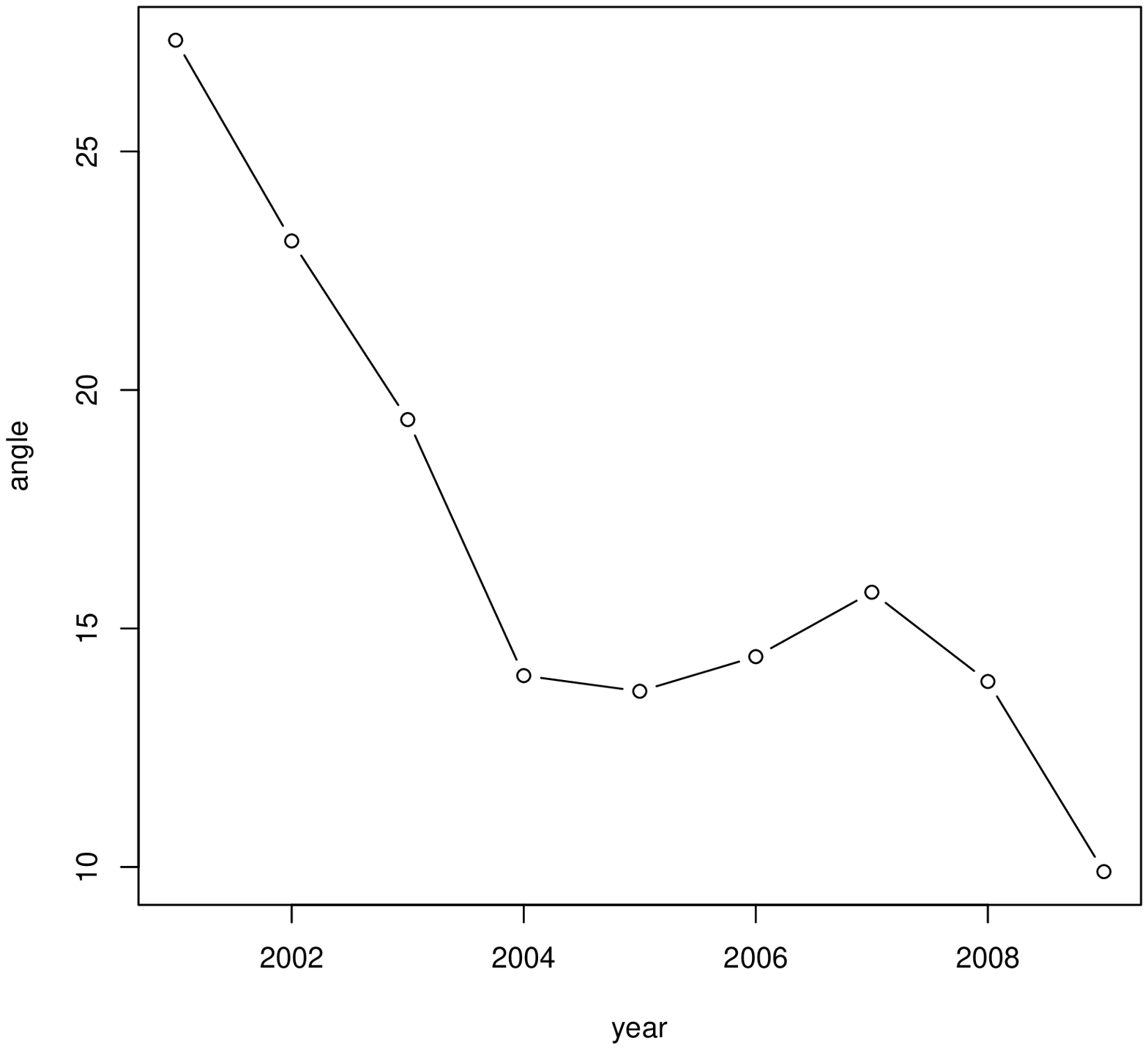}%
\end{minipage}
\end{figure}

\begin{figure}
\caption{\label{fig:sparsecom} Heatmap of the support of the error component
by LOREC from 2001-2009.}

\begin{centering}
\includegraphics[width=0.8\textwidth]{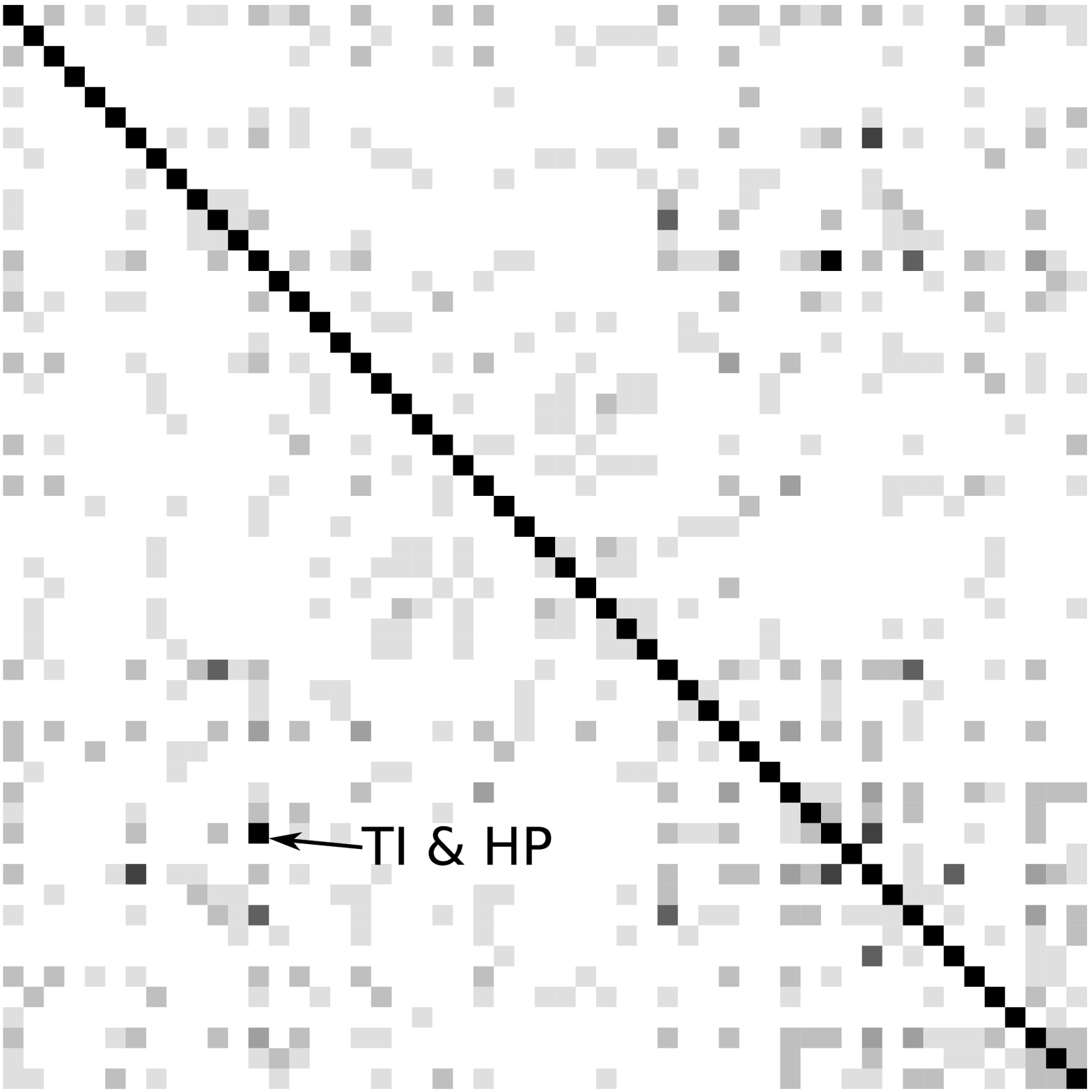}
\par\end{centering}

\end{figure}

Besides the low rank component, LOREC also recovers a sparse component
for each year, and the heatmap of support recovery through 2001-2009
is plotted in Figure \ref{fig:sparsecom}. There are some interesting
patterns revealed by LOREC. For instance, a nonzero correlation between
Texas Instrument (TI) and Hewlett Packard (HP) is identified through
all these years. A possible explanation is that they are both involved
in the business of semiconductor hardware, and they can be influenced
by the same type of news (fluctuations). Other significant nonzero
correlations are also examined, and they seem to be reasonable within
this type of explanation.

\section{Discussion\label{sec:discuss} }

In this paper, we propose a new covariance structure framework that
recovers the covariance structures from many popular statistical models.
The statistical performance is theoretically justified using various
losses. We develop a first-order algorithm and prove the iteration
complexity bounds of this algorithm. The merits of this new approach
are illustrated using numerical examples. 

One future direction is to further understand other theoretical properties
under different norms, especially the minimax optimal rates. The Frobenius
norm result here is conjectured to be minimax optimal using a similar
argument from \citet{RohdeA11}, but it remains to be rigorously justified.
Moreover, the convergence rates are shown to be different under the
Frobenius norm and the operator norm \citep{CaiT10a}, and it is interesting
to study if such a phenomenon also exists under the covariance structure
\eqref{eq:lowspvar}. 

The results of course can be further improved with stronger structural
assumptions. For example, we consider the factors to be unobservable.
If one assumes that reasonable proxies are available, it is an interesting
problem to incorporate such additional information, and improvements
may be achieved hence. Moreover, the conditions imposed are to derive
both support/rank recovery and matrix loss bounds. If one is interested
in the joint matrix losses, without separating the two components,
we suspect a different set of conditions shall suffice, for example
\citet{AgarwalA2011}. Different conditions should also apply if one
is only interested in determining the number of factors. Finally,
it is also interesting to explore other extensions of general framework.
For example, we will report the multiple spiked covariance model elsewhere. 

The tuning parameters can be chosen theoretically as in the main results,
or chosen using cross validation in practice. It is of great interest
to study data-driven choices. We found that the development in \citet{JohnstoneI09}
is particularly promising. On a related topic of estimating the inverse covariance matrix, we
established the theoretical justification for using cross validation
\citep{LiuLuo2012}. In our low rank plus sparse setting, the analysis for data-driven
penalties is still an open problem. 

The convex optimization approach is adopted here to illustrate the
framework, and for the computational efficiency reason. It is interesting
to study adaptive versions of our proposal to ameliorate the bias.
For example, one may replace the penalty function $\pen\left(\cdot\right)$
in \eqref{eq:lsobj} by either the adaptive lasso penalty \citep{ZouH06}
or the SCAD penalty \citep{FanJ01}. We leave this to future work.
It is also interesting to explore constrained versions of LOREC, because
there were some improvements when estimating the inverse covariance
\citep{CaiT2010}.

Our algorithm may benefit further from incorporating additional numerical
tricks, in order to improve the computation speed empirically. Currently
full SVD is used for updating the low rank component. Partial SVD,
such as partial Lanczos SVD, should suffice. In particular, it may
provide additional numerical advantages for large scale problems.
We plan to implement this in future releases of our package.

\bibliographystyle{apa}
\bibliography{../../rossi}

\begin{thebibliography}{}

\bibitem[\protect\astroncite{Agarwal et~al.}{2011}]{AgarwalA2011}
Agarwal, A., Negahban, S., and Wainwright, M. (2011).
\newblock Noisy matrix decomposition via convex relaxation: Optimal rates in
  high dimensions.
\newblock {\em arXiv preprint arXiv:1102.4807}.

\bibitem[\protect\astroncite{Amini and Wainwright}{2009}]{AminiA09}
Amini, A.~A. and Wainwright, M.~J. (2009).
\newblock High-dimensional analysis of semidefinite relaxations for sparse
  principal components.
\newblock {\em Annals of Statistics}, 37:2877--2921.

\bibitem[\protect\astroncite{Anderson}{1984}]{AndersonT84}
Anderson, T.~W. (1984).
\newblock {\em An Introduction to Multivariate Statistical Analysis}.
\newblock John Wiley, New York.

\bibitem[\protect\astroncite{Beck and Teboulle}{2009}]{BeckA09}
Beck, A. and Teboulle, M. (2009).
\newblock {A fast iterative shrinkage-thresholding algorithm for linear inverse
  problems}.
\newblock {\em SIAM Journal on Imaging Sciences}, 2(1):183--202.

\bibitem[\protect\astroncite{Bickel and Levina}{2008a}]{BickelP08}
Bickel, P. and Levina, E. (2008a).
\newblock {Regularized estimation of large covariance matrices}.
\newblock {\em Annals of Statistics}, 36(1):199--227.

\bibitem[\protect\astroncite{Bickel and Levina}{2008b}]{BickelP08b}
Bickel, P.~J. and Levina, E. (2008b).
\newblock Covariance regularization by thresholding.
\newblock {\em Annals of Statistics}, 36(6):2577--2604.

\bibitem[\protect\astroncite{Bunea et~al.}{2011}]{BuneaF11}
Bunea, F., She, Y., and Wegkamp, M.~H. (2011).
\newblock Optimal selection of reduced rank estimators of high-dimensional
  matrices.
\newblock {\em The Annals of Statistics}, 39(2):1282--1309.

\bibitem[\protect\astroncite{Cai et~al.}{2010a}]{CaiJ10}
Cai, J.-F., Candes, E.~J., and Shen, Z. (2010a).
\newblock A singular value thresholding algorithm for matrix completion.
\newblock {\em SIAM J Optim}, 20:1956--1982.

\bibitem[\protect\astroncite{Cai and Liu}{2011}]{CaiT2011}
Cai, T. and Liu, W. (2011).
\newblock Adaptive thresholding for sparse covariance matrix estimation.
\newblock {\em Journal of the American Statistical Association, to appear.},
  106(494):672--684.

\bibitem[\protect\astroncite{Cai et~al.}{2011}]{CaiT2010}
Cai, T., Liu, W., and Luo, X. (2011).
\newblock A constrained $\ell_1$ minimization approach to sparse precision
  matrix estimation.
\newblock {\em Journal of the American Statistical Association},
  106(494):594--607.

\bibitem[\protect\astroncite{Cai et~al.}{2010b}]{CaiT10a}
Cai, T., Zhang, C., and Zhou, H. (2010b).
\newblock {Optimal rates of convergence for covariance matrix estimation}.
\newblock {\em Ann. Statist}, 38:2118--2144.

\bibitem[\protect\astroncite{Candes et~al.}{2009}]{CandesE2009}
Candes, E., Li, X., Ma, Y., and Wright, J. (2009).
\newblock {Robust principal component analysis}.
\newblock {\em preprint submitted to Journal of the ACM}.

\bibitem[\protect\astroncite{Chamberlain and Rothschild}{1983}]{ChamberlainG83}
Chamberlain, G. and Rothschild, M. (1983).
\newblock {Arbitrage, factor structure, and mean-variance analysis on large
  asset markets}.
\newblock {\em Econometrica: Journal of the Econometric Society},
  51(5):1281--1304.

\bibitem[\protect\astroncite{Chandrasekaran et~al.}{2010}]{ChandrasekaranV2010}
Chandrasekaran, V., Parrilo, P., and Willsky, A. (2010).
\newblock {Latent Variable Graphical Model Selection via Convex Optimization}.
\newblock {\em Arxiv preprint arXiv:1008.1290}.

\bibitem[\protect\astroncite{Davis and Kahan}{1970}]{DavisC70}
Davis, C. and Kahan, W. (1970).
\newblock The rotation of eigenvectors by a perturbation. iii.
\newblock {\em SIAM Journal on Numerical Analysis}, 7(1):1--46.

\bibitem[\protect\astroncite{Fama and French}{1992}]{FamaE92}
Fama, E. and French, K. (1992).
\newblock {The cross-section of expected stock returns}.
\newblock {\em Journal of finance}, 47(2):427--465.

\bibitem[\protect\astroncite{Fan et~al.}{2008}]{FanJ08}
Fan, J., Fan, Y., and Lv, J. (2008).
\newblock High dimensional covariance matrix estimation using a factor model.
\newblock {\em Journal of Econometrics}, 147(1):186 -- 197.

\bibitem[\protect\astroncite{Fan and Li}{2001}]{FanJ01}
Fan, J. and Li, R. (2001).
\newblock Variable selection via nonconcave penalized likelihood and its oracle
  properties.
\newblock {\em J. Amer. Statist. Assoc.}, 96:1348--1360.

\bibitem[\protect\astroncite{Fazel et~al.}{2001}]{FazelM01}
Fazel, M., Hindi, H., and Boyd, S. (2001).
\newblock A rank minimization heuristic with application to minimum order
  system approximation.
\newblock In {\em American Control Conference}, volume~6, pages 4734 --4739
  vol.6.

\bibitem[\protect\astroncite{Fitzmaurice et~al.}{2004}]{Fitzmaurice04}
Fitzmaurice, G., Laird, N., and Ware, J. (2004).
\newblock {\em {Applied longitudinal analysis}}.
\newblock Wiley-Interscience.

\bibitem[\protect\astroncite{Foster and George}{1994}]{FosterD94}
Foster, D.~P. and George, E.~I. (1994).
\newblock The risk inflation criterion for multiple regression.
\newblock {\em The Annals of Statistics}, 22(4):1947--1975.

\bibitem[\protect\astroncite{Friedman et~al.}{2008}]{FriedmanJ08}
Friedman, J., Hastie, T., and Tibshirani, R. (2008).
\newblock Sparse inverse covariance estimation with the graphical lasso.
\newblock {\em Biostatistics}, 9(3):432--441.

\bibitem[\protect\astroncite{Horn and Johnson}{1994}]{HornR94}
Horn, R. and Johnson, C. (1994).
\newblock {\em {Topics in matrix analysis}}.
\newblock Cambridge Univ Pr.

\bibitem[\protect\astroncite{Huang et~al.}{2006}]{HuangJ06}
Huang, J., Liu, N., Pourahmadi, M., and Liu, L. (2006).
\newblock Covariance matrix selection and estimation via penalised normal
  likelihood.
\newblock {\em Biometrika}, 93:85--98.

\bibitem[\protect\astroncite{Jennrich and Schluchter}{1986}]{Jennrich86}
Jennrich, R. and Schluchter, M. (1986).
\newblock {Unbalanced repeated-measures models with structured covariance
  matrices}.
\newblock {\em Biometrics}, 42(4):805--820.

\bibitem[\protect\astroncite{Ji and Ye}{2009}]{JiS09}
Ji, S. and Ye, J. (2009).
\newblock {An accelerated gradient method for trace norm minimization}.
\newblock In {\em Proceedings of the 26th Annual International Conference on
  Machine Learning}, pages 457--464. ACM.

\bibitem[\protect\astroncite{Johnstone and Lu}{2009}]{JohnstoneI09}
Johnstone, I. and Lu, A. (2009).
\newblock {On consistency and sparsity for principal components analysis in
  high dimensions}.
\newblock {\em Journal of the American Statistical Association},
  104(486):682--693.

\bibitem[\protect\astroncite{Koltchinskii et~al.}{2011}]{KoltchinskiiV11}
Koltchinskii, V., Lounici, K., and Tsybakov, A.~B. (2011).
\newblock Nuclear-norm penalization and optimal rates for noisy low-rank matrix
  completion.
\newblock {\em The Annals of Statistics}, 39(5):2302--2329.

\bibitem[\protect\astroncite{Krim and Viberg}{1996}]{KrimH96}
Krim, H. and Viberg, M. (1996).
\newblock Two decades of array signal processing research: the parametric
  approach.
\newblock {\em Signal Processing Magazine, IEEE}, 13(4):67 --94.

\bibitem[\protect\astroncite{Ledoit and Wolf}{2003}]{LedoitO03}
Ledoit, O. and Wolf, M. (2003).
\newblock {Improved estimation of the covariance matrix of stock returns with
  an application to portfolio selection}.
\newblock {\em Journal of Empirical Finance}, 10(5):603--621.

\bibitem[\protect\astroncite{Ledoit and Wolf}{2004}]{LedoitO04}
Ledoit, O. and Wolf, M. (2004).
\newblock A well-conditioned estimator for large-dimensional covariance
  matrices.
\newblock {\em Journal of Multivariate Analysis}, 88(2):365 -- 411.

\bibitem[\protect\astroncite{Leek and Storey}{2007}]{LeekJ07}
Leek, J.~T. and Storey, J.~D. (2007).
\newblock Capturing heterogeneity in gene expression studies by surrogate
  variable analysis.
\newblock {\em PLoS Genetics}, 3(9):e161.

\bibitem[\protect\astroncite{Liu and Luo}{2012}]{LiuLuo2012}
Liu, W. and Luo, X. (2012).
\newblock High-dimensional sparse precision matrix estimation via sparse column
  inverse operator.
\newblock {\em arXiv preprint arXiv:1203.3896}.

\bibitem[\protect\astroncite{Markowitz}{1952}]{MarkowitzH52}
Markowitz, H. (1952).
\newblock {Portfolio selection}.
\newblock {\em Journal of Finance}, 7(1):77--91.

\bibitem[\protect\astroncite{Meinshausen and
  B\"{u}hlmann}{2006}]{MeinshausenN06}
Meinshausen, N. and B\"{u}hlmann, P. (2006).
\newblock High-dimensional graphs and variable selection with the lasso.
\newblock {\em The Annals of Statistics}, 34(3):1436--1462.

\bibitem[\protect\astroncite{Nesterov}{1983}]{NesterovY83}
Nesterov, Y. (1983).
\newblock {A method of solving a convex programming problem with convergence
  rate O (1/k2)}.
\newblock In {\em Soviet Mathematics Doklady}, volume~27, pages 372--376.

\bibitem[\protect\astroncite{Nesterov and Nesterov}{2004}]{NesterovY04}
Nesterov, Y. and Nesterov, I. (2004).
\newblock {\em {Introductory lectures on convex optimization: A basic course}}.
\newblock Kluwer Academic Publishers.

\bibitem[\protect\astroncite{Onatski}{2009}]{OnatskiA09}
Onatski, A. (2009).
\newblock Testing hypotheses about the number of factors in large factor
  models.
\newblock {\em Econometrica}, 77(5):1447--1479.

\bibitem[\protect\astroncite{Rohde and Tsybakov}{2011}]{RohdeA11}
Rohde, A. and Tsybakov, A.~B. (2011).
\newblock Estimation of high-dimensional low-rank matrices.
\newblock {\em The Annals of Statistics}, 39(2):887--930.

\bibitem[\protect\astroncite{Ross}{1976}]{RossS76}
Ross, S. (1976).
\newblock {The arbitrage theory of capital asset pricing}.
\newblock {\em Journal of Economic Theory}, 13:341--360.

\bibitem[\protect\astroncite{Rothman et~al.}{2009}]{RothmanA09}
Rothman, A.~J., Levina, E., and Zhu, J. (2009).
\newblock Generalized thresholding of large covariance matrices.
\newblock {\em Journal of the American Statistical Association},
  104(485):177--186.

\bibitem[\protect\astroncite{Sharpe}{1964}]{SharpeW64}
Sharpe, W. (1964).
\newblock {Capital asset prices: A theory of market equilibrium under
  conditions of risk}.
\newblock {\em The Journal of Finance}, 19(3):425--442.

\bibitem[\protect\astroncite{Tibshirani}{1996}]{TibshiraniR96}
Tibshirani, R. (1996).
\newblock Regression shrinkage and selection via the lasso.
\newblock {\em Journal of the Royal Statistical Society. Series B
  (Methodological)}, 58(1):267--288.

\bibitem[\protect\astroncite{Wu and Pourahmadi}{2003}]{WuW03}
Wu, W.~B. and Pourahmadi, M. (2003).
\newblock Nonparametric estimation of large covariance matrices of longitudinal
  data.
\newblock {\em Biometrika}, 90(4):831--844.

\bibitem[\protect\astroncite{Yuan}{2010}]{YuanM10}
Yuan, M. (2010).
\newblock High dimensional inverse covariance matrix estimation via linear
  programming.
\newblock {\em Journal of Machine Learning Research}, 11:2261--2286.

\bibitem[\protect\astroncite{Zou}{2006}]{ZouH06}
Zou, H. (2006).
\newblock The adaptive lasso and its oracle properties.
\newblock {\em Journal of the American Statistical Association},
  101:1418--1429.

\end{thebibliography}

\end{document}